\documentclass[a4paper,11pt]{article}
\pdfoutput=1 

\usepackage{jheppub} 

\usepackage{color}
\usepackage{braket}
\usepackage{ytableau}
\usepackage[vcentermath]{youngtab}

\usepackage[T1]{fontenc} 

\title{\boldmath Interacting Emergent Yang-Mills theory}

\author[a]{Chenliang Su}

\affiliation[a]{Guangdong Provincial Key Laboratory of Nuclear Science, Institute of Quantum Matter,\\South China Normal University, Higher Education Mega Center,\\West Waihuan Road No. 378, Guangzhou, China}

\emailAdd{suchenliang@m.scnu.edu.cn}

\abstract{In this article, subleading (in $1/N$) corrections to the action of the one loop dilatation operator in the su(3) sector of $\mathcal{N}=4$ super Yang-Mills theory are studied. We focus on the system of operators dual to two giant graviton systems, which have a bare dimension $\sim\mathcal{O}(N)$ and are a linear combination of restricted Schur polynomials with $p=2$ long columns. At the leading order the dilatation operator gives rise to the free part of an emergent Yang-Mills theory, arising from the open string excitations of the giant gravitons. We verify that the terms we study describe interactions between these open string excitations. The interactions have the U(1)$\times$U(1) gauge invariance expected for a pair of separated branes.}

\begin{document} 
	\maketitle
	\flushbottom
	
	\section{Introduction}
	
	The AdS/CFT correspondence \cite{1999TheLargeN,1998Gaugetheory,witten1998anti} claims an exact duality between ${\cal N}=4$ super Yang-Mills theory and string theory on spacetimes that are asymptotically AdS$_5\times$S$^5$. The correspondence can be used to show that states in the string theory correspond to operators in the Yang-Mills theory, and the converse. In this article we are interested in operators that have a bare dimension of order $N$ and correspond to giant graviton branes in the dual string theory \cite{2000Invasion,2000SUSY,2000Largebranes}. Giant gravitons are spherical branes that carry a D3-brane dipole charge. As usual, to construct excited D-brane states, we attach open strings to the brane. At low energy these open strings should be described by a Yang-Mills theory, so that the low energy dynamics of these operators should be described by an emergent Yang-Mill theory\cite{2005DbranesInYangMills}. Our goal is to study the one loop dilatation operator acting on the operators dual to giant graviton branes, in order to test this idea. This was performed in the leading large $N$ limit in \cite{2020EmergentYangMills}. The result matches a free emergent Yang-Mills theory associated with the brane world volume dynamics. Our goal in this paper is to compute the first ${1\over N}$ correction, in order to learn about interactions.
    
	
	The Study on the action of the dilatation operator, at large $N$, for operators with a dimension of order $N$ is highly nontrivial \cite{2002GiantGravitons}. The large $N$ limit is usually dominated by planar diagrams, and higher genus ribbon graphs are suppressed\cite{1974Aplanardiagram}. However, for operators of dimension of order $N$, the combinatorics of the Feynman diagrams can be used to show that the sheer number of non-planar diagrams overpowers the higher genus suppression. The usual simplifications for large $N$ do not hold and new ideas are needed. We will follow the approach to this problem based on representation theory, developed in \cite{corley2002exact,2007GiantgravitonsI,2008ExactMultiMatrix,2007BranesAntiBranes,2008DiagonalMultiMatrix,2009DiagonalFreeField,2008EnhancedSymmetries,2008ExactMultiRestrictedSchur}. These works developed bases spanning the space of the local gauge invariant operators of the model, that diagonalize two point functions in the free field theory exactly (i.e. to all orders in ${1\over N}$), while at low loop order these operators mix only weakly . Specifically, we will use the basis provided by the restricted Schur polynomials \cite{2008ExactMultiMatrix}.
	
	
	Our approach entails evaluating the exact (to all orders in ${1\over N}$) action for the one loop dilatation operator and then expanding to extract the leading term and the first subleading correction. Even with the powerful representation theory methods, this is a problem of considerable complexity, so that we will focus on a system of two giant gravitons. This corresponds to studying restricted Schur polynomials labeled by Young diagrams with two long columns. Since it plays a central role in our analysis, we briefly review the derivation of how the dilatation operator acts on the restricted Schur polynomials in Section \ref{Dact}. The leading large $N$ dilatation operator can be diagonalized analytically, with the eigenstates known as ``Guass graph operators''. The construction of the Gauss graph operators is reviewed in section 
	\ref{GGBasis}. In Section \ref{emergentYM} we explain the identification of the Gauss graph operators with states in the Hilbert space of an emergent Yang-Mills theory. The discussion at this point has all dealt with the leading contribution in a ${1\over N}$ expansion, which reproduces the free emergent Yang-Mills theory. Having set the stage, we are now ready to turn to evaluate the first subleading corrections, which represent interactions in the emergent Yang-Mills. To do this we calculate the exact action of the one loop dilatation operator in Section \ref{analyticsubleading}. Our computation involves three complex scalars and so is a generalization of similar computations presented in \cite{2010Emergentthree,2011Surprisingly}. This generalization is necessary as the operator mixing corresponding to interactions in the emergent Yang-Mills is not captured by the study in \cite{2010Emergentthree,2011Surprisingly}. Our final result for the interaction is in Section \ref{sbldin}. This is the key result of this paper. Conclusions and discussion are given in Section \ref{discuss}.
	
	The physics of how open strings and their dynamics emerges from ${\cal N}=4$ super Yang-Mills theory is a fascinating subject and there are by now many papers on this topic. We recommend \cite{2005QuantizingOpenSpin,2013OpenSpinChains,2015GiantGravitonsAndTheEmergence,berenstein2015central,2015ExcitedStates,berenstein2020open,2021OpenGiantMagnons,2022arXiv220211729B} and their references, for background.

	\section{Action of the Dilatation Operator in Restricted Schur Polynomials basis}\label{Dact}
	
	In this section we review the exact\footnote{i.e. to all orders in $1/N$.} action of the one loop dilatation operator on restricted Schur polynomials in the $SU(3)$ sector of $\mathcal{N}=4$ super Yang-Mills theory. The one loop dilatation operator \cite{2003TheBetheAnsatz} in this case reads
	\begin{equation}
		D=-\frac{2g_{YM}^2}{(4\pi)^2}\left([\phi_3,\phi_1][\partial_{\phi_{3}},\partial_{\phi_{1}}]+[\phi_{2},\phi_{1}][\partial_{\phi_{2}},\partial_{\phi_{1}}]+[\phi_{3},\phi_{2}][\partial_{\phi_{3}},\partial_{\phi_{2}}]\right)
	\end{equation}
	For convenience, we introduce the notation ($A,B=1,2,3$)
	\begin{equation}
		D\equiv	-\frac{2g_{YM}^2}{(4\pi)^2} \sum_{A>B=1}^{3} D_{AB}
	\end{equation}
	where $D_{AB}$ mixes fields $\phi_A$ and $\phi_B$. We will consider the action of this dilatation operator on the restricted Schur polynomials\cite{2013JHEP...03..173D}
	\begin{equation}
		\begin{aligned}
			\chi_{R,(\vec{r})\vec{\mu}\vec{\nu}}\left(\phi\right)&=\frac{1}{n_1 ! n_2 ! n_3 !}\sum_{\sigma \in S_{n_T}}\chi_{R(\vec{r})\vec{\mu}\vec{\nu}}(\sigma) \operatorname{Tr}\left(\sigma \phi_1^{\otimes n_1} \phi_2^{\otimes n_2} \phi_3^{\otimes n_3}  \right)
		\end{aligned}
	\end{equation}
	where $n_T=n_1+n_2+n_3$ and
	\begin{equation}
		\begin{aligned}
			\operatorname{Tr}\left(\sigma \phi_1^{\otimes n_1} \phi_2^{\otimes n_2} \phi_3^{\otimes n_3}  \right)=(\phi_1)_{i_{\sigma(1)}}^{i_1}\cdots (\phi_1)_{i_{\sigma(n_1)}}^{i_{n_1}}(\phi_2)_{i_{\sigma(n_1+1)}}^{i_{n_1+1}}\cdots (\phi_2)_{i_{\sigma(n_1+n_2)}}^{i_{n_1+n_2}}\\
			\times (\phi_3)_{i_{\sigma(n_1+n_2+1)}}^{i_{n_1+n_2+1}}\cdots (\phi_3)_{i_{\sigma(n_1+n_2+n_3)}}^{i_{n_1+n_2+n_3}}
		\end{aligned}
	\end{equation}
	The restricted Schur polynomials are labeled by representations and multiplicity labels. $R$ denotes an irreducible representation of $S_{n_{T}}$, labeled by a Young diagram with $n_T$ boxes. We have $\vec{r}=(r_1,r_2,r_3)$, where $r_A$ is a Young diagram with $n_A$ boxes, and $n_1+n_2+n_3=n_T$. Together these three Young diagrams label an irreducible representation of $S_{n_1}\times S_{n_2}\times S_{n_3}$ which is a subgroup of $S_{n_{T}}$. We know the operator should be invariant under swapping bosons $\phi_A$. The only way to realize this is to make the row and column indices of $\phi_A$ transform in the identical representation $r_A$ and then project onto an uniquely trivial representation in $r_A\otimes r_A$. Further we have $\vec{\mu}=(\mu_2,\mu_3)$ and $\vec{\nu}=(\nu_1,\nu_2)$, where $\mu_A$ and $\nu_A$ specifies the multiplicity of $r_A$ as a subspace of the carrier space of $R$. \ At this point we are forced to introduce multiplicities because the representation of the subgroup may appear more than once. We remove $n_2$ boxes from $R$, and assemble them into $r_2$. There might be more than one way to do this, while they bring us into different copies of the carrier space of $r_2$. We distinguish different copies using $\mu_2$ (or $\nu_2$). Similarly, removing $n_3$ boxes from $R$ we might find different copies of $r_3$, and we use $\mu_3$ (or $\nu_3$) to distinguish them. Finally the remaining boxes in $R$ compose the Young diagram labeling $r_1$ so that no multiplicity label is needed for it.
	
	$\chi_{R,(\vec{r})\vec{\mu}\vec{\nu}}(\sigma)$ is a restricted character\cite{2007Giantgravitons}, obtained by summing over the row index of $\Gamma^{R}(\sigma)$ over the subspace $(\vec{r})\vec{\mu}$ and the column index over the subspace $(\vec{r})\vec{\nu}$ which both arise upon restricting $R$ of $S_{n_{T}}$ to its $S_{n_1}\times S_{n_2}\times S_{n_3}$ subgroup, as explained above. It is useful to write
	\begin{equation}
		\chi_{R,(\vec{r}) \vec{\mu} \vec{\nu}}(\sigma)=\operatorname{Tr}_{R}\left(P_{R,(\vec{r}) \vec{\mu} \vec{\nu}} \Gamma^{(R)}(\sigma)\right)
	\end{equation}	
	The trace is over the carrier space of irreducible representation $R$. The operator $P_{R,(\vec{r})\vec{\mu}\vec{\nu}}$ is an intertwining map. In the above trace, it makes the row indices of $\Gamma^{R}(\sigma)$ over the copy of $(\vec{r})$ labeled by $\vec{\nu}$ and the column indices over the copy of $\vec{r}$ labeled by $\vec{\mu}$. 
	
	For convenience we will use the restricted Schur polynomials normalized to have a unit two point function. The normalized operator $O_{R,(\vec{r})\vec{\mu}\vec{\nu}}\left(\phi\right)$ is defined by	
	\begin{equation}
		\chi_{R,(\vec{r})\vec{\mu}\vec{\nu}}\left(\sigma\right)=\sqrt{\frac{f_{R}\text{hooks}_{R}}{\prod_{A}\text{hooks}_{r_A}}}O_{R,(\vec{r})\vec{\mu}\vec{\nu}}\left(\sigma\right)
	\end{equation}

	We will show the action of dilatation operator on normalized restricted Schur polynomials in what follows. It is useful to introduce the short hand
	\begin{equation}
		\begin{array}{ll}
			1_{\phi_{1}}=n_2+n_3+1 & n_{\phi_{1}}=n_1+n_2+n_3=n_{T} \\
			1_{\phi_{2}}=n_{1}+1 & n_{\phi_{2}}=n_{1}+n_{2} \\
			1_{\phi_{3}}=1 & n_{\phi_{3}}=n_{1} \\
		\end{array}
	\end{equation}
	We also simplify $1_{\phi_{A}}$ as $1_{A}$. Note that $n_{\phi_A}$ and $n_{A}$ are distinct and should not be confused. Now the action of the dilatation operator\cite{2020EmergentYangMills} reads
	\begin{equation}
		\begin{aligned}
			D O_{R(\vec{r})\vec{\mu}\vec{\nu}}=-\frac{2g^2_{YM}}{(4\pi)^2}\sum_{A>B=1}^{3}\sum_{T(\vec{t})\vec{\alpha}\vec{\beta}}(\mathcal{M}_{AB})_{R(\vec{r})\vec{\mu}\vec{\nu},T(\vec{t})\vec{\alpha}\vec{\beta}}O_{T(\vec{t})\vec{\beta}\vec{\alpha}}
		\end{aligned}\label{dilone}
	\end{equation}
	\begin{equation}\label{action on restricted Schur's}
		\begin{aligned}
			(\mathcal{M}_{AB})_{R(\vec{r})\vec{\mu}\vec{\nu},T(\vec{t})\vec{\alpha}\vec{\beta}}&=\sum_{R',T'}\sqrt{c_{RR'}c_{TT'}}\sqrt{\frac{\text{hooks}_{\vec{r}} \text{hooks}_{\vec{t}}}{\text{hooks}_{R} \text{hooks}_{T}}}\frac{n_A n_B \sqrt{\text{hooks}_{R'} \text{hooks}_{T'}}}{n_1!n_2!n_3!}\\
			\times&\text{Tr}_{R}\Big( \left[ \Gamma^R((1,1_A))P_{R(\vec{r})\vec{\mu}\vec{\nu}}\Gamma^R((1,1_A)),\Gamma^R((1,1_B))\right] I_{R'T'} \\
			&\times  \left[\Gamma^T((1,1_A))P_{T(\vec{t})\vec{\alpha}\vec{\beta}}^{\dagger}\Gamma^T((1,1_A)),\Gamma^T((1,1_B))\right] I_{T'R'}\Big)
		\end{aligned}
	\end{equation}
	where $R'$ and $T'$ are irreducible representations of $S_{n_T-1}$ obtained by removing one box from $R$ and $T$ respectively. $I_{R'T'}$ is an interwiner from $T'$ to $R'$, i.e. we have $I_{R'T'}=0$ if $R'\neq T'$. One can refer to \cite{2020EmergentYangMills} for a detailed derivation.
	
	They are the expressions defined by (\ref{dilone}) and (\ref{action on restricted Schur's}) that will in the end be used to define the Hamiltonian of the emergent gauge theory. To make the connection we need to study the dilatation operator in a basis that makes the connection to excited brane states most transparent. This basis, known as the Gauss graph basis, is introduced in the next section.
	
	\section{Diagonalization in the Gauss Graph Basis}\label{GGBasis}
	
	In this section we will diagonalize the dilatation operator in its $(\vec{r})\vec{\mu}$ indices, by moving to the Gauss graph basis. We will give a brief introduction to the Gauss graph basis. This involves defining the displaced corners limit at large $N$. Finally, we will prove that the conclusions given in \cite{2020EmergentYangMills}, obtained in the long rows case, also follow in the long columns case which we are considering. This is the first new result of this paper.
	
	\subsection{Transforming to Gauss graph basis}
	
	It is useful to begin with a motivation for the Gauss graph basis. We consider operators which have a definite semi-classical limit in the holographically dual theory, which allows us to simplify the large $N$ dynamics. In this paper we consider a system of $p$ giant gravitons, dual to operators labeled by Young diagram $R$ with $p$ long columns. We call this the long columns case, while the system of operators labeled by Young diagram with long rows is called the long rows case. The long rows case has been discussed in \cite{2020EmergentYangMills}. 
	We study the operators with a dimension $\Delta \sim N$, so that there should be $\sim N$ boxes in the Young diagram $R$ that labels the operator. To construct these operators, many $\phi_1$ fields and a few $\phi_2,\phi_3$ fields as excitations are used. Precisely, we assume $n_1 \sim N$ and $n_2\sim n_3\sim \sqrt{N}$.
	
	These operators mix with each other only if the Young diagrams labeling them own the same amount of rows or columns. It has been argued that corners on the bottom of the Young diagram $R$ are well separated at large $N$ and weak coupling\cite{2020EmergentYangMills}. This is called the displaced corners limit, which simplifies the the action of the symmetric group on boxes at the corners: permutations just swap boxes they act on. Noticing that to obtain irreducible representations of the subgroup we remove and then reassemble boxes at corners, this simplication implies new symmetries and conservation laws \cite{2011GiantGravitonOscillators}. The new symmetry is swapping  row or column indices of $\phi_A$ that belong to the same column. The new conservation law is that operators mix only if the numbers of boxes removed from each column to obtained the Young diagram $r_A$ are the same. This fact motivates the notation $\vec{n}_{A}=((n_{A})_{1},(n_{A})_{2},\dots,(n_{A})_{p})$ where $(n_A)_{i}$ tells us how many boxes are removed from the $i$th column of $R$ and then assembled into $r_A$. Using this notation the group representing the new symmetry is
	\begin{equation}
		H_{\vec{n}_{A}}=S_{(n_{A})_{1}}\times S_{(n_{A})_{2}}\times\cdots \times S_{(n_{A})_{p}}
	\end{equation}
	Both row and column indices of $\phi_A$ fields have this symmetry. Thus inequivalent operators constructed from the $\phi_{A}$ fields are specified by elements of the double coset
	\begin{equation}
		H_{\vec{n}_{A}}\backslash S_{n_{A}} /H_{\vec{n}_{A}}
	\end{equation}
	Since the above double coset contains the same number of elements as that of triples $(r_A,\mu_A,\nu_A)$, we are allowed to organize $\phi_A$ fields using the elements of this double coset instead of the triple $(r_A,\mu_A,\nu_A)$\cite{2012Adoublecosetansatz}. In what follows we show the relevant double cosets we will use to label our operators
	\begin{equation}
		\begin{aligned}
			&\phi_2 \leftrightarrow \sigma_{2} \in H_{\vec{n}_{2}}\backslash S_{n_{2}} /H_{\vec{n}_{2}}\\
			&\phi_3 \leftrightarrow \sigma_{3} \in H_{\vec{n}_{3}}\backslash S_{n_{3}} /H_{\vec{n}_{3}}
		\end{aligned}
	\end{equation}
	where we use $\sigma_A$ to refer to an element of the double coset $H_{\vec{n}_{A}}\backslash S_{n_{A}} /H_{\vec{n}_{A}}$. For convenience we will use the notation $\vec{\sigma}=(\sigma_2,\sigma_{3})$. It is clear that $\vec{\sigma}$ refers to an element of a direct product of two double cosets.
	
	The Gauss graph provides a graphical description of $\vec{\sigma}$. A Gauss graph consists of distinguishable nodes and directed edges stretching between nodes. We allow an edge to return to where it departs, 
	but at each node, the numbers of edges departing and arriving should be equal. This constraint follows from the Gauss law of the emergent gauge theory \cite{2005DbranesInYangMills,2012Adoublecosetansatz}. More details of the connection between graphs and elements of a double coset can be found in \cite{2012StringsfromFeynman}. It is evident that both permuting edges departing from a given node and permuting those arriving at a given node yields an identical Gauss graph. Thus, non-equivalent Gauss graphs can be specified by elements of a double coset and serve as a graphical description of them. In our cases, a Gauss graph describing $\vec{\sigma}$ has $p$ nodes corresponding to columns of Young diagram $R$. There is a species of edges related to each type of $\phi_2$ and $\phi_3$ fields, while the number of edges of this species is respectively determined by the number of fields, $n_2$ and $n_3$. For each $\vec{\sigma}$, the relevant Gauss graph shows a specified configuration. An example of the configuration of the Gauss graph is shown in Figure 1 of \cite{2012Adoublecosetansatz}. 
	To describe a Gauss graph we let $(n_{A})_{i\rightarrow j}$ denote the number of edges stretching from node $i$ to node $j$, while $(n_{A})_{ij}=(n_{A})_{i\rightarrow j}+(n_{A})_{j\rightarrow i}$ denotes the amount of edges connecting node $i$ and node $j$. In particular, we assume $(n_{A})_{ii}=(n_{A})_{i\rightarrow i}$.
	
	Following \cite{2020EmergentYangMills} we will transform from the restricted Schur polynomial basis to the Gauss graph basis. Since we are considering the displaced corners limit when $R$ has $p$ long columns, some modification of the discussion of \cite{2020EmergentYangMills} is needed. We will use the group theoretical coefficients 
	\begin{equation}\label{group theoretical coefficients for long columns}
		C^{r_A}_{\mu_{A} \nu_{A}}(\tau)=\left|H_{\vec{n}_A}\right| \sqrt{\frac{d_{r_A}}{n_A !}}\sum_{k,l=1}^{d_{r_A}} \Gamma^{(r_A^T)}_{k l}(\tau) B_{k \mu_{A}}^{r_A \rightarrow 1^{\vec{n}_A}} B_{l \nu_{A}}^{r_A \rightarrow 1^{\vec{n}_A}}
	\end{equation}
	to transform the labels of $\phi_{A}$ fields, where $\tau$ is an element of $S_{n_{A}}$, $d_{r_{A}}$ is the dimension of $r_{A}$, $\left|H_{\vec{n}_A}\right|$ is the order of $H_{\vec{n}_A}$, and $\Gamma^{(r_A^T)}_{k l}(\tau)$ is the matrix representing $\tau$ in the representation $r_A^T$, which is the conjugate representation of $r_A$. $B_{k \mu_{A}}^{r_A \rightarrow 1^{\vec{n}_A}}$ is a branching coefficient, defined by
	\begin{equation}
		\sum_{\mu_A} B_{k \mu_{A}}^{r_A \rightarrow 1^{\vec{n}_A}} B_{l \mu_{A}}^{r_A \rightarrow 1^{\vec{n}_A}}=\frac{1}{|H_{\vec{n}_{A}}|}\sum_{\gamma\in H_{\vec{n}_{A}}}\Gamma^{(r_A^T)}_{kl}(\gamma)
	\end{equation}
	where $1^{\vec{n}_A}$ denotes the anti-trivial representation of $H_{\vec{n}_A}$, which might appear more than once in $r_A$. $\mu_A$ labels these multiple copies so that $\mu_A$ runs from 1 to the number of copies of $1^{\vec{n}_A}$ in $r_A$. 
	One might be concerned that $\mu_A$ has already been used to specify the multiplicity of $r_A$ as a subspace of the carrier space of $R$. We use this notation on purpose since it has been proved in \cite{2012Adoublecosetansatz} that the number of copies of $r_A$ in $R$ is equal to that of $1^{\vec{n}_A}$ in $r_A$, if we remove $(n_A)_{i}$ boxes from the $i$th column of $R$ to obtain $r_A$.
	Using these coefficients, we define the Gauss graph operators by
	\begin{equation}\label{basis transformation}
		O_{R,r_1}(\vec{\sigma})=\sum_{r_2\vdash n_2}\sum_{r_3\vdash n_3}\sum_{\vec{\mu},\vec{\nu}}C^{r_2}_{\mu_{2} \nu_{2}}(\sigma_2)C^{r_3}_{\mu_{3} \nu_{3}}(\sigma_3)O_{R,(\vec{r})\vec{\mu}\vec{\nu}}
	\end{equation}
	Thus, we need a Gauss graph and two Young diagrams $R$ as well as $r_1$ to label a Gauss graph operator, while this Gauss graph, as discussed above, describes element $\vec{\sigma}$ of a direct product of double cosets. For the sake of an unit two point function, we then define the normalized operator $\hat{O}_{R,r_1}(\vec{\sigma})$ by
	\begin{equation}
		O_{R,r_1}(\vec{\sigma})=\sqrt{\prod_{A=2}^{3}\prod_{i,j=1}^{p}(n_A)_{i\rightarrow j}!}\hat{O}_{R,r_1}(\vec{\sigma})
	\end{equation}
	
	We now turn to the action of the dilatation operator in the Gauss graph basis. To obtain it we need to evaluate
	\begin{equation}\label{transform to Gauss graph basis}
		(M_{AB})_{R,r_1,\vec{\sigma}; T,t_1,\vec{\tau}} =
		\sum_{\substack{
				r_2,r_3,\vec{\mu},\vec{\nu}\\
				t_2,t_3,\vec{\alpha},\vec{\beta}}}
		C^{(r_2,r_3)}_{\vec{\mu}\vec{\nu}}(\vec{\sigma}) C^{(t_2,t_3)}_{\vec{\alpha}\vec{\beta}}(\vec{\tau})(\mathcal{M}_{AB})_{R(\vec{r})\vec{\mu}\vec{\nu},T(\vec{t})\vec{\alpha}\vec{\beta}}
	\end{equation}
	where
	\begin{equation}
		C^{(r_2,r_3)}_{\vec{\mu}\vec{\nu}}(\vec{\sigma})=C^{r_2}_{\mu_{2} \nu_{2}}(\sigma_2)C^{r_3}_{\mu_{3} \nu_{3}}(\sigma_3)
	\end{equation}
	The detailed calculation is given in \cite{2020EmergentYangMills}. Our case is almost identical so we will simply quote the result
	\begin{equation}\label{D31,D21 in Gauss graph basis}
		\begin{aligned}
			&D_{31}O_{R,r_1}(\vec{\sigma})=\sum_{i>j=1}^{p}(n_3)_{ij}\Delta_{ij}O_{R,r_1}(\vec{\sigma})\\
			&D_{21}O_{R,r_1}(\vec{\sigma})=\sum_{i>j=1}^{p}(n_2)_{ij}\Delta_{ij}O_{R,r_1}(\vec{\sigma})\\
		\end{aligned}
	\end{equation}
	where operator $\Delta_{ij}$ acts only on the $R,r_1$ labels and we have $\Delta_{ij}=\Delta^{0}+\Delta^{+}+\Delta^{-}$. Denote the length of the $i$th column of the Young diagram $r$ by $l_{i}$. Young diagram $r^{+}_{ij}$ is obtained by removing a box from column $j$ and adding it to column $i$. Young diagram $r^{-}_{ij}$ is obtained by removing a box from column $i$ and adding it to column $j$. 
	Using this notation we can write the action of $\Delta^{0}$ and $\Delta^{\pm}$ as
	\begin{equation}
		\begin{aligned}
			&\Delta^{0}_{ij}O_{R,r}(\vec{\sigma})=-(2N-l_{r_i}-l_{r_j})O_{R,r}(\vec{\sigma})\\
			&\Delta^{\pm}_{ij}O_{R,r}(\vec{\sigma})=\sqrt{(N-l_{r_i})(N-l_{r_j})}O_{R^{\pm}_{ij},r^{\pm}_{ij}}(\vec{\sigma})\\
		\end{aligned}
	\end{equation}
	The action of $D_{32}$, which is described by matrix $M_{32}$, is more complicated. It is given by
	\begin{equation}\label{M32 in Gauss graph basis}
		\begin{aligned}
			&\left(M_{3 2}\right)_{R, r_{1}, \vec{\sigma}; T, t_{1}, \vec{\tau}}= \sum_{R^{\prime}_{i},T'_{j}} \frac{\delta_{r_{1} t_{1}} \delta_{R_{i}^{\prime} T_{j}^{\prime}}}{\sqrt{\left|O_{R, r_{1}}\left(\vec{\sigma}\right)\right|^{2}\left|O_{T, t_{1}}\left(\vec{\tau}\right)\right|^{2}}} \sqrt{\frac{c_{R R_{i}^{\prime}} c_{T T_{j}^{\prime}}}{l_{R_{i}} l_{T_{j}}}} \\
			&\times\left[ 2\delta_{ik}(n_3)_{i}(n_2)_{i}-\left( (n_3)_{ki}(n_2)_{ii}+(n_{3})_{ii}(n_2)_{ik} \right) \right]\sum_{\gamma_1,\gamma_2\in H_2}\delta(\vec{\sigma}^{-1}\gamma_1\vec{\tau}\gamma_2)
		\end{aligned}
	\end{equation}
	where $R'_{i}$ denotes the Young diagram produced by removing one box from the $i$th column, and $c_{R R_{i}^{\prime}}$ is the factor of the removed box. We assume that $t_A$ is obtained by removing $(n'_A)_i$ boxes from the $i$th column of $T$ and $H_2=H_{\vec{n}'_3}\times H_{\vec{n}_2} $. 
	A Gauss graph has the norm as what follows
	\begin{equation}
		\left| O_{R,r}(\vec{\sigma}) \right|^2=\prod_{i,j=1}^{p}\left(n_{2}^{\vec{\sigma}}\right)_{i\rightarrow j}!\left(n_{3}^{\vec{\sigma}}\right)_{i\rightarrow j}!
	\end{equation}
	where the superscript $\vec{\sigma}$ of $n_A$ indicates it is counting edges in the Gauss graph describing $\vec{\sigma}$. In this paper, we focus on the matrix $M_{32}$ which describes the interaction between excitations.
	
	\subsection{Long Columns}
	
	Our discussion has frequently used results obtained in \cite{2020EmergentYangMills}, in which the discussion about the displaced corner limit is given in the case that restricted Schur polynomials have long rows. But the same conclusions follow for the long column case, and we will explain it in this section.
	
	Firstly, we argue that the action of the dilatation operator on restricted Schur polynomials, $(\mathcal{M}_{AB})_{R(\vec{r})\vec{\mu}\vec{\nu},T(\vec{t})\vec{\alpha}\vec{\beta}}$, in the long column case has the same form as that in the long row case. 
	In the displaced corners limit, swapping boxes in the same column yields a minus sign. Thus, $\Gamma^R((1,1_A))$ should be identified with $\text{sgn}((1,1_A))(1,1_A)=-(1,1_A)$, while $\Gamma^R((1,1_B))$ should be identified with $-(1,1_B)$. In equation (\ref{action on restricted Schur's}) we find four $\Gamma^R((1,1_A))$ and two $\Gamma^R((1,1_B))$, which gives rise to $(-1)^6=1$. Thus, this action is identical in both cases of the displaced corners limit.
	
	Next, the action of the dilatation operator on Gauss graph operators, $(M_{AB})_{R,r_1,\vec{\sigma}; T,t_1,\vec{\tau}}$, will also be proved identical in both cases. 
	Note that we obtain this action by using group theoretical coefficients $C^{r_A}_{\mu_A \nu_A}$ to perform the transformation shown in equation (\ref{transform to Gauss graph basis}). The coefficient used in the long row case reads
	\begin{equation}
		\widetilde{C}^{r_A}_{\mu_{A} \nu_{A}}(\tau)=\left|H_{\vec{n}_A}\right| \sqrt{\frac{d_{r_A}}{n_A !}}\sum_{k,l=1}^{d_{r_A}} \Gamma^{(r_A)}_{k l}(\tau) B_{k \mu_{A}}^{r_A \rightarrow 1_{H_{\vec{n}_A}}} B_{l \nu_{A}}^{r_A \rightarrow 1_{H_{\vec{n}_A}}}
	\end{equation}
	where $1_{H_{\vec{n}_A}}$ denotes the trivial representation of $H_{\vec{n}_A}$ as a subspace of $r_A$. The tilde, as in $\widetilde{C}^{r_A}_{\mu_{A} \nu_{A}}(\tau)$ is used to distinguish this branching coefficient from the coefficient used in the long column case. The equality
	\begin{equation}\label{equality about branching coefficients}
		\sum_{\mu}B_{i\mu}^{r\rightarrow 1^{\vec{m}}} B_{j\mu}^{r\rightarrow 1^{\vec{m}}}=\sum_{\mu}B_{i\mu}^{r^T\rightarrow 1_{H_{\vec{m}}}} B_{j\mu}^{r^T\rightarrow 1_{H_{\vec{m}}}}=\frac{1}{|H_{\vec{m}}|}\sum_{\gamma\in H_{\vec{m}}}\Gamma_{ij}^{(r^T)}(\gamma)
	\end{equation}
	is useful in what follows. In the detailed calculation of the change, all branching coefficients are all summed in this way. Clearly, if we replace $C^{r_A}_{\mu_{A} \nu_{A}}(\tau)$ with $\widetilde{C}^{r_A}_{\mu_{A} \nu_{A}}(\tau)$ and $\sum_{r_A\vdash n_A}$ with $\sum_{r_A^T\vdash n_A}$ in the transformation (\ref{transform to Gauss graph basis}), the final result is unchanged, thanks to the equality (\ref{equality about branching coefficients}). This proves that the action of the dilatation operator in Gauss graph basis given in the last section is the correct result in the long column case.

	\section{Emergent Yang-Mills Theory: at The Leading Order}\label{emergentYM}
	
	In this section, we will interpret the action of the dilatation operator at the leading order, arguing that gives the dynamics of a Yang-Mills theory. Basically, we firstly identify the action shown in equations (\ref{D31,D21 in Gauss graph basis}) and (\ref{M32 in Gauss graph basis}) with the Hamiltonian describing the dynamics of some states written using an occupation number representation. Each state corresponds to a Gauss graph. Then, this Hamiltonian is recognized as the Hamiltonian of the worldvolume dynamics of giant graviton branes, which is a super Yang-Mills theory.

	\subsection{Emergent Hamiltonian}
	
	The action of the dilatation operator $D_{32}$ on Gauss graph operators, given by equation $(\ref{M32 in Gauss graph basis})$, can naturally be identified as a Hamiltonian of oscillators. It is evident that $D_{32}$ acts only on Gauss graph labels so that operators don't mix if their $R,r_1$ labels are different. Further, the configuration of the Gauss graph can be interpreted as an occupation number representation of the state by identifying an oriented edge for $\phi_A$ field stretching from the $i$th node to $j$th node as a particle created by the creation operator $(\bar{b}_A)_{ij}$. With these states, it is possible to rewrite $D_{32}$ in terms of some creation and annihilation operators. They are divided into two categories, one corresponding to each type of field. 
	
	We perform this rewriting next. Introduce the oscillators $(\bar{b}_2)_{ij},(b_2)_{ij}$ to describe $\phi_2$ field, as well as oscillators $(\bar{b}_3)_{ij},(b_3)_{ij}$ for the $\phi_3$ field. The oscillator algebra reads
	\begin{equation}
		[(b_A)_{ij},(\bar{b}_B)_{kl}]=\delta_{AB}\delta_{il}\delta_{jk}; \quad [(b_A)_{ij},(b_A)_{kl}]=[(\bar{b}_A)_{ij},(\bar{b}_B)_{kl}]=0
	\end{equation} 
	where $A,B=2,3$. We can also define the number operators by
	\begin{equation}
		(\hat{n}_A)_{i\rightarrow j}=(\bar{b}_A)_{ij}(b_A)_{ji}; \quad
	\end{equation}
	it is convenient to introduce  $(\hat{n}_A)_{ij}=(\hat{n}_A)_{ji}=(\hat{n}_A)_{i\rightarrow j}+(\hat{n}_A)_{j\rightarrow i}$ (in particular, $(\hat{n}_A)_{ii}=(\hat{n}_A)_{i\rightarrow i}$), and 
	\begin{equation}
		(\hat{n}_A)_{i}=\sum_{j=1}^p (\hat{n}_A)_{ij}=\sum_{j=1}^{p} (\hat{n}_A)_{ji}.
	\end{equation}
	A state corresponding to the Gauss graph operator $O_{R,r}(\vec{\sigma})$ is given by
	\begin{equation}
		O_{R,r}(\vec{\sigma}) \leftrightarrow \prod_{i,j=1}^{p} (\bar{b}_2)_{ij}^{(n_2^{\vec{\sigma}})_{i\rightarrow j}}(\bar{b}_3)_{ij}^{(n_3^{\vec{\sigma}})_{i\rightarrow j}}\ket{0}
	\end{equation}
	where the vaccum state $\ket{0}$ obeys $(a_{A})_{ij}\ket{0}=0$ for any $A,B=2,3$ and $i,j=1,2,\dots,p$. The $R,r_1$ labels seem to be missing on the RHS, but, as we discuss above, the dynamics between states with the same $R,r_1$ labels completely describe the action of $D_{32}$, so that we can drop them to simplify our discussion. With this identification, the action of $D_{32}$ can be rewritten as\cite{2020EmergentYangMills}
	\begin{equation}
		\begin{aligned}
			H_{32}=\sum_{i,j=1}^{p}&\sqrt{\frac{(N-l_{r_i})(N-l_{r_j})}{l_{r_i} l_{r_j}}}\\
			&\times\Big( -(\hat{n}_3)_{ji}(\bar{b}_2)_{jj}(b_2)_{ii}-(\hat{n}_2)_{ji}(\bar{b}_3)_{jj}(b_3)_{ii}+2\delta_{ij}(\hat{n}_{3})_{i}(\hat{n}_{2})_{i} \Big)
		\end{aligned}
	\end{equation}
	Note that only the closed loop at nodes in the Gauss graph are dynamical---they are the only objects that will be annihilated or created.
	
	Similarly, we can simplify the action of $D_{31}$ and $D_{21}$, shown in equation (\ref{D31,D21 in Gauss graph basis}), at large $N$ and further identify them with terms $H_{31}$ and $H_{21}$ in the Hamiltonian. These terms describe the interaction between the states introduced above. At large $N$, we identify $O_{R_{ij}^{\pm},r_{ij}^{\pm}}(\vec{\sigma})$ with $O_{R,r}(\vec{\sigma})$ so that the $R,r_1$ labels are again not needed to describe these terms. After simplification, the action of $D_{31}$ and $D_{21}$ can be rewritten as
	\begin{equation}
		\begin{aligned}
			&H_{31}=-\sum_{i,j=1}^{p}\left(\sqrt{N-l_{r_i}}-\sqrt{N-l_{r_j}}\right)^2(\hat{n}_3)_{i\rightarrow j}\\
			&H_{21}=-\sum_{i,j=1}^{p}\left(\sqrt{N-l_{r_i}}-\sqrt{N-l_{r_j}}\right)^2(\hat{n}_2)_{i\rightarrow j}\\
		\end{aligned}
	\end{equation}
	The complete Hamiltonian, identified with the action of the dilatation operator, is $H=H_{32}+H_{31}+H_{21}$. We will soon see that this Hamiltonian describes the dynamics of a Yang-Mills theory.
	
	\subsection{Identification with a Yang-Mills Theory}
	
	We will prove that the emergent Hamiltonian obtained in the last subsection matches with the Hamiltonian of a Yang-Mills theory, which we call the emergent Yang-Mills theory. What Yang-Mills theory do we expect? The operators we study have $p$ long columns so that this system can be identified with a system of $p$ giant gravitons based on the holographic duality. It motivates us to guess that the emergent Hamiltonian describes the worldvolume dynamics of these gravitons. We expect this world volume dynamics, which should be a super Yang-Mills theory since it arises from the dynamics of open string excitations stretching between the giant graviton branes, to match with the emergent Hamiltonian we derive.
	
	To confirm our expectation, we will explicitly write down the Hamiltonian of this Yang-Mills theory for comparison. We expect a $U(p)$ gauge theory as the system consists of $p$ giant graviton branes. Each column in Young diagram $R$, or equivalently, each node in the Gauss graph corresponds to a brane. Therefore, the edges are naturally interpreted as open string excitations. The branes move in AdS$_5$ spacetime with metric
	\begin{equation}
		ds^2=R^2\left( -\cosh^2\rho dt^2 + d\rho^2+ \sinh^2\rho d\Omega_{3}^{2} \right)
	\end{equation}
	A brane corresponding to the $i$th column in $R$ has $\rho$ coordinate specified by
	\begin{equation}
		\cosh\rho =\sqrt{1+\frac{l_{R_i}}{N}} \qquad \sinh\rho=\sqrt{\frac{l_{R_i}}{N}}
	\end{equation}
	In the displaced corners limit $l_{R_i}$ are not equal, hence the branes are separated, which implies that the Coulomb branch of this gauge theory is being considered. In the low energy limit the dynamics is described by a $U(1)^p$ gauge theory. In addition, since the $\text{su}(3)$ sector is part of the $\text{su}(2|3)$ sector of the $\mathcal{N}=4$ super Yang-Mills theory, we do not expect to the recover the complete $U(1)^p$ gauge theory. In fact, we should reproduce part of the $s-$wave bosonic sector of the emergent Yang-Mills theory \cite{2020EmergentYangMills}.
	
	The above discussion motivates us to study a $U(p)$ gauge theory of adjoint scalars living on an $S^3$. The action reads
	\begin{equation}
		\begin{gathered}
			S=\frac{1}{g_{Y M}^{2}} \int_{\mathbb{R} \times S^{3}}\left[\operatorname{Tr}\left(\partial_{\mu} X \partial^{\mu} X^{\dagger}+\partial_{\mu} Y \partial^{\mu} Y^{\dagger}-\frac{1}{R^{2}}\left(X X^{\dagger}+Y Y^{\dagger}\right)-[X, Y][Y^{\dagger}, X^{\dagger}]\right)\right. \\
			\left.-\sum_{i \neq j} m_{i j}^{2}\left(X_{i j} X_{j i}^{\dagger}+Y_{i j} Y_{j i}^{\dagger}\right)\right] d t R^{3} d \Omega_{3}
		\end{gathered}
	\end{equation}
	where $m_{ij}$ are masses of the off diagonal matrix elements of $X,Y$, proportional to the distances separating the branes. The $s-$wave sector is given by
	\begin{equation}
		\begin{aligned}
			S=\frac{R^{3} \Omega_{3}}{g_{Y M}^{2}} \int_{\mathbb{R}} &\left[\operatorname{Tr}\left(\dot{X} \dot{X}^{\dagger}+\dot{Y} \dot{Y}^{\dagger}-\frac{1}{R^{2}}\left(X X^{\dagger}+Y Y^{\dagger}\right)-[X, Y][Y^{\dagger}, X^{\dagger}]\right)\right.\\
			&\left.-\sum_{i \neq j} m_{i j}^{2}\left(X_{i j} X_{j i}^{\dagger}+Y_{i j} Y_{j i}^{\dagger}\right) d t\right]
		\end{aligned}
	\end{equation}
	The action of the one-loop dilatation operator should correspond to the interaction Hamiltonian give by
	\begin{equation}
		H_{\text{int}}=\frac{R^{3} \Omega_{3}}{g_{Y M}^{2}}\left[ \sum_{i\neq j}^{p}m_{ij}^{2}\left(X_{i j} X_{j i}^{\dagger}+Y_{i j} Y_{j i}^{\dagger}\right)+\text{Tr}\left([X, Y][Y^{\dagger}, X^{\dagger}]\right) \right]
	\end{equation} 
	Since our operators in the original theory are constructed using $\phi_A$ but not $\phi_A^{\dagger}$, a proper truncation should be made in the emergent theory. This truncation is acheieved by setting $X=\bar{a},X^{\dagger}=a,Y=\bar{b}$ and $Y^{\dagger}=b$. Then, the truncated normal ordered interaction Hamiltonian is given by
	\begin{equation}
		H_{\text{int}}=\frac{R^{3} \Omega_{3}}{g_{Y M}^{2}}\left[ \sum_{i\neq j}^{p}m_{ij}^{2}\left(\bar{a}_{i j} a_{j i}+\bar{b}_{i j} b_{j i}\right)+\text{Tr}\left([\bar{b}, \bar{a}][a, b]\right) \right]
	\end{equation} 
	By identifying $\bar{b},b$ with $\bar{b}_3,b_3$ and $\bar{a},a$ with $\bar{b}_2,b_2$, we can prove the identification in what follows
	\begin{equation}
		\renewcommand{\arraystretch}{1.5}
		\begin{array}{lll}
			m_{ij}^{2}\bar{a}_{i j} a_{j i} &\leftrightarrow & H_{31}\\
			m_{ij}^{2}\bar{b}_{i j} b_{j i}&\leftrightarrow& H_{21}\\
			\text{Tr}\left([\bar{b}, \bar{a}][a, b]\right) &\leftrightarrow& H_{32}
		\end{array}
	\end{equation} 
	A detailed discussion is given in \cite{2020EmergentYangMills}. Now we have confirmed that at the leading order, the action of the dilatation operator perfectly describes the dynamics of giant graviton branes.

	\section{The Subleading Corrections}\label{analyticsubleading}
	
	In this section, we will focus on a system of two giant gravitons, in which the Young diagram labeling the operator has two long columns. We also assume that $n_2=n_3=2$, i.e. our operators are constructed using many $\phi_1$ fields, two $\phi_2$ fields and two $\phi_3$ fields. This restriction significantly simplify our formulas so that an analytic calculation of matrix elements of the dilatation operator is possible. We will perform this calculation in this section, obtaining an explicit expression for the action of the dilatation operator in the Gauss graph basis, which is valid to all orders in $1/N$. This exact result will be expanded to obtain the subleading correction to the leading action. The leading action has already been matched to Yang-Mills theory at linear level. The subleading corrections correspond to interactions. Finally, we find the spectrum of the leading contribution to $D_{32}$ has evenly spaced energy levels, described by a formula which fixes the size of the energy to be an order 1 number at large $N$, while the subleading corrections cause a small correction to the leading energy level.

	\subsection{The System of Two Giant Gravitons}
	
	We study a set of operators labeled by Young diagram $R$ with only two columns. The operators are constructed using many $\phi_1$ fields, two $\phi_2$ fields and two $\phi_3$ fields, i.e. we set $p=2$ and $n_3=n_2=2$ and take $n_1$ to be of order $N$. These choices imply some significant simplifications:
	\begin{itemize}
		\item[1.] Although the dilatation operator acting on operators labeled by Young diagrams with two columns produces operators with three columns, the contribution of these three column operators will be neglected---the extra column of these operators is much shorter than the two long columns, so they correspond to bound states of two threebranes with some Kaluza-Klein (KK) gravitons. And graviton emission happens when a two threebrane state transforms into a state of two threebranes with KK gravitons, which implies that in the ’t Hooft limit the amplitude of this transition is proportional to the string coupling $g_s \propto \frac{1}{N}$. Thus, the mixing with more than two columns operators is suppressed and can be dropped. This simplification allows us to study the action of the dilatation operator in the basis consisting of a finite number of operators, each with two long columns. Precisely, the restricted Schur polynomials basis consists of 16 types of operators as shown below
		\begin{equation}\label{restricted Schur's basis of two giant gravitons}
			\begin{aligned}
				&O_{aa}(\xi_0,\xi_1);\quad O_{da}(\xi_0,\xi_1);\quad O_{ea}(\xi_0,\xi_1);\quad O_{ba}(\xi_0,\xi_1)\\
				&O_{ad}(\xi_0,\xi_1);\quad O_{ae}(\xi_0,\xi_1);\quad O_{dd}(\xi_0,\xi_1);\quad O_{de}(\xi_0,\xi_1)\\
				&O_{ed}(\xi_0,\xi_1);\quad O_{ee}(\xi_0,\xi_1);\quad O_{bd}(\xi_0,\xi_1);\quad O_{be}(\xi_0,\xi_1)\\
				&O_{ab}(\xi_0,\xi_1);\quad O_{db}(\xi_0,\xi_1);\quad O_{eb}(\xi_0,\xi_1);\quad O_{bb}(\xi_0,\xi_1)\\
			\end{aligned}
		\end{equation}
		Our notation is as follows: we use $(\xi_0,\xi_1)$ to specify the Young diagram $r_1$ which has $\xi_0$ rows with two boxes and $\xi_1$ rows with one box. $a,b,c,d$ are used to specify the the irreducible representation $r_A$ and $\vec{n}_A$, the vector with elements as amounts of boxes removed from each column of the Young diagram $R$ to be assembled into $r_A$, as what follows
		\begin{equation}
			\begin{aligned}
				&a \rightarrow {\tiny \young({\,}{\,},{\,}{\,},{\,}{\,},{\,}{\,},{\,},{*},{*})}; {\tiny \yng(2,2,2,2,1)},{\tiny \yng(1,1)}\\
				&b \rightarrow {\tiny \young({\,}{\,},{\,}{\,},{\,}{*},{\,}{*},{\,},{\,},{\,})}; {\tiny \yng(2,2,1,1,1,1,1)},{\tiny \yng(1,1)}\\
				&d \rightarrow {\tiny \young({\,}{\,},{\,}{\,},{\,}{\,},{\,}{*},{\,},{\,},{*})}; {\tiny \yng(2,2,2,1,1,1)},{\tiny \yng(2)}\\
				&e \rightarrow {\tiny \young({\,}{\,},{\,}{\,},{\,}{\,},{\,}{*},{\,},{\,},{*})}; {\tiny \yng(2,2,2,1,1,1)},{\tiny \yng(1,1)}\\
			\end{aligned}
		\end{equation}
		where on the RHS the first Young diagram denotes $R$ while the last Young diagram denotes $r_A$. We label the boxes to be removed by $*$. For example, by removing one box from each column of $R$ and assembling them into ${\tiny \yng(2)}$ we obtain $d$, so that we have $\vec{n}=(1,1)$. Both $r_2$ and $r_3$ can be one of $a,b,c,d$ so that we have $4\times4=16$ operators, while the first subscript specifies $r_2$ and the second subscript specifies $r_3$. For instance, $O_{ae}(\xi_0,\xi_1)$ denote the operator labeled by $r_1=(\xi_0,\xi_1),r_2=a,r_3=e$ with $\vec{n}_2=(2,0),\vec{n}_3=(1,1)$---we first remove one box from each column of $R$ and reassemble them into ${\tiny \yng(1,1)}$ to obtain $r_3=e$, and then remove two boxes from the first column of $R$ to obtain $r_2=a$. Note that when deciding how removed boxes can be assembled to give an irreducible representation, we must respect edges that are joined. The remaining part of $R$ is nothing but $r_1=(\xi_0,\xi_1)$.
		
		\item[2.] It can be proved that when the Young diagram $R$ has only two columns, no multiplicity label $\vec{\mu}$ is needed, i.e. each $\vec{r}=(r_1,r_2,r_3)$, the irreducible representation of $S_{n_1}\times S_{n_2}\times S_{n_3}$ as the subgroup of $S_{n_T}$, appear only once as the subspace of $R$ \cite{2011Surprisingly}. This fact simplifies the interwiner $P_{R,(\vec{r}) \vec{\mu} \vec{\nu}}$ as the projector $P_{R\rightarrow(\vec{r})}$ from $R$ onto its subspace $\vec{r}$. It allows us to explicitly write down $P_{R\rightarrow(\vec{r})}$ using the basis of $\vec{r}$ represented by the Young-Yamanouchi symbol.
		
		As an example, we will explain how to obtain $P_{R\rightarrow(\xi_0,\xi_1)da}$, where the subspace it projects onto is specified by $r_1=(\xi_0,\xi_1),r_2=d,r_3=a$. This subspace should be spanned by
		\begin{equation}
			\ket{1,i}=\ket{{\tiny \young({\,}{\,},{\,}{\,},{\,}{\,},{\,}{4},{\,},{3},{2},{1})}\,i};\quad \ket{2,i}=\ket{{\tiny \young({\,}{\,},{\,}{\,},{\,}{\,},{\,}{3},{\,},{4},{2},{1})}\,i}
		\end{equation}
		where we assume $S_{n_3}$ acts on numbers $1,2$ while $S_{n_2}$ acts on $3,4$. We use label $i$ to denote $d_{r_1}$ different ways to fill out the remaining boxes. Thus, the above basis spans a $2d_{r_1}-$dimensional space given by $(\xi_0,\xi_1)da\oplus (\xi_0,\xi_1)ea$. The irreducible representation $(\xi_0,\xi_1)da$ is furnished by vectors 
		\begin{equation}
			\begin{aligned}
				&\ket{{\tiny \young({\,}{\,},{\,}{\,},{\,}{\,},{\,},{\,})}\,i,{\tiny \young({4}{3})},{\tiny \young({2},{1})}}=A\ket{1,i}+B\ket{2,i}\\
			\end{aligned}
		\end{equation}
		Allowing $\Gamma^{R}((34))$ to act on the above vector and using the well-known action of $\Gamma^{R}(\sigma)$ on the Young-Yamanouchi vector, we obtain the following equation
		\begin{equation}
			\begin{aligned}
				&A=-\frac{1}{(\xi_1+2)}A+\sqrt{1-\frac{1}{(\xi_1+2)^2}}B\\
				&B=\sqrt{1-\frac{1}{(\xi_1+2)^2}}A+\frac{1}{(\xi_1+2)}B
			\end{aligned}
		\end{equation}
		With the normalization $A^2+B^2=1$, we obtain a solution
		\begin{equation}
			A=\sqrt{\frac{\xi_1+1}{2(\xi_1+2)}}; \quad B=\sqrt{\frac{\xi_1+3}{2(\xi_1+2)}};
		\end{equation} 
		Finally, we can write the projector $P_{R\rightarrow(\xi_0,\xi_1)da}$ as
		\begin{equation}
			\begin{aligned}
				P_{R\rightarrow(\xi_0,\xi_1)da}&=\sum_{i} \ket{{\tiny \young({\,}{\,},{\,}{\,},{\,}{\,},{\,},{\,})}\,i,{\tiny \young({4}{3})},{\tiny \young({2},{1})}}\bra{{\tiny \young({\,}{\,},{\,}{\,},{\,}{\,},{\,},{\,})}\,i,{\tiny \young({4}{3})},{\tiny \young({2},{1})}}\\
				&=\sum_i \left(\frac{\xi_1+1}{2(\xi_1+2)}\ket{1,i}\bra{1,i} + \frac{\sqrt{(\xi_1+1)(\xi_1+3)}}{2(\xi_1+2)}\ket{1,i}\bra{2,i} \right. \\
				&\left. + \frac{\sqrt{(\xi_1+1)(\xi_1+3)}}{2(\xi_1+2)}\ket{2,i}\bra{1,i} + \frac{\xi_1+3}{2(\xi_1+2)}\ket{2,i}\bra{2,i} \right)
			\end{aligned}
		\end{equation}
		
		\item[3.] Labeling the Young diagram $r_1$ by a pair of parameters $(\xi_0,\xi_1)$ makes it straight forward to consider the large $N$ limit and the displaced corners limit. We assume $\xi_0\sim N$ and $\xi_0>>\xi_1$ when we refer to the large $N$ limit, hence we are allowed to set $\xi_0+\xi_1+a=\xi_0$ where $a$ is any number of order 1. When we refer to the displaced corners limit, we assume $\xi_1\sim \sqrt{N}$, which allows us to set $\xi_1+a=\xi_1$ and drop terms of order higher than (or equal to) $\frac{1}{\xi_1}$. Our computation thus has two small numbers, $\frac{1}{\xi_0}$ and $\frac{1}{\xi_1}$. We can expand in these two numbers, dropping higher order terms which provides a dramatic simplification.
		
	\end{itemize}
	
	Thanks to this simplification, it is possible to perform an analytic calculation to derive the exact action of the dilatation operator. We will explain this calculation next.
	
	\subsection{The Analytic Calculation}
	
	Our goal is to obtain the action of the dilatation operator in the Gauss graph basis, using the displaced corners limit, which is expected to describe the worldvolumne dynamics of giant gravitons. Based on the discussion in above sections, our calculation can be summarized as
	\begin{itemize}
		\item[1] Calculate the action of the dilatation operator in the restricted Schur polynomial basis according to equation (\ref{action on restricted Schur's}). The bulk of the work for this step is to compute the trace $\text{Tr}_{R}(\cdots)$ by expressing operators traced in terms of the Young-Yamanouchi vectors, which give a basis for $R$. In this step, we end up obtaining some $16\times16$ matrices describing the action of $D_{31},D_{21},D_{32}$ in the restricted Schur polynomial basis. 
		\item[2] Perform the transformation shown in equation (\ref{transform to Gauss graph basis}) on the matrices obtained in the last step. This entails computing the group theoretic coefficients given in equation (\ref{group theoretical coefficients for long columns}). This gives three $16\times16$ matrices, describing the action of $D_{31},D_{21},D_{32}$ in the Gauss graph basis. These matrices have entries dependent on $\xi_0,\xi_1$.
		\item[3] Finally, we can expand the answer, after performing a power series expansion in the two small parameters $1/\xi_0$ and $1/\xi_1$. In order to refer to the large $N$ limit, we first expand results in $1/\xi_0$ and retain terms of the leading order of it. In practice, terms of order $(1/\xi_0)^0$ are retained in $D_{31}$ and $D_{21}$, while those of order $1/\xi_0$ are retained in $D_{32}$. Then, we expand answers in $1/\xi_1$. Eventually, the leading contribution to $D_{31}$ and $D_{21}$ comes from terms of order $1/(\xi_0\xi_1)^0$, and the leading contribution to $D_{32}$ is given by terms of order $1/(\xi_0\xi_1^{\,0})$. Therefore, the subleading correction to $D_{32}$ is given by terms of order $1/(\xi_0\xi_1)$. The leading order must agree with the analytic results shown in equations (\ref{D31,D21 in Gauss graph basis}) and $(\ref{M32 in Gauss graph basis})$, which provides a highly non-trivial check of our analytic result. The subleading terms correspond to interactions. These are new results obtained for the first time in this paper and they allow us to study the non-linear interactions present in the emergent Yang-Mills theory. 
	\end{itemize}
	
	It is worth describing some of the details needed for the calculation. Firstly, we will explain how to compute the trace $\text{Tr}_{R}(\cdots)$ in equation (\ref{action on restricted Schur's}). It is principally done by writing everything in terms of the Young-Yamanouchi vectors. As the action of $\Gamma^{R}(\sigma)$ in the Young-Yamanouchi basis is widely known and we know how to write down the projectors that were introduced above, we need only discuss the interwiners $I_{R'T'}$. This map is non-vanishing only when $R$ is related to $T$, either $(i)$ by removing one box from the second column of $T$ to the first column, $(ii)$ by removing one box from the first column of $T$ to the second column or (iii) if $R=T$. The $I_{R'T'}$ for case $(i)$ can be written as
	\begin{equation}
		I_{R'T'}=\ket{{\tiny \young({\,}{\,},{\,}{\,},{\,}{\,},{\,}{\,},{\,},{\,},{\,},{1})}}\bra{{\tiny \young({\,}{\,},{\,}{\,},{\,}{\,},{\,}{\,},{\,}{1},{\,},{\,})}}
	\end{equation}
	while $I_{T'R'}$ is manifestly obtained by swapping the bra and ket vectors
	\begin{equation}
		I_{T'R'}=\ket{{\tiny \young({\,}{\,},{\,}{\,},{\,}{\,},{\,}{\,},{\,}{1},{\,},{\,})}}\bra{{\tiny \young({\,}{\,},{\,}{\,},{\,}{\,},{\,}{\,},{\,},{\,},{\,},{1})}}
	\end{equation} 
	The interwiners for cases $(ii)$ and $(iii)$ are written in a similar way.
	
	Next, we will derive the group theoretical coefficients used to transform to the Gauss graph basis. The main task is to compute the branching coefficient, which is given by
	\begin{equation}
		B_{k\mu_A}^{r_A\rightarrow 1^{\vec{n}_A}}=\braket{r_A;k|r\rightarrow 1^{\vec{n}_A};\mu_A}
	\end{equation}
	where $\ket{r_A;k}$ denotes the $k$th basis vector of $r_A$ and $\ket{r\rightarrow 1^{\vec{n}_A};\mu_A}$ denote the basis vector of anti-trivial representation $1^{\vec{n}_A}$ as a 1-dimensional subspace of $r_A$. In our two giant graviton system, no multiplicity label is needed and all possible $r_A$, given by $a,b,c,d$ introduced above, are one-dimensional. Thus we can drop all subscripts on $B_{k\mu_A}^{r_A\rightarrow 1^{\vec{n}_A}}$ and write it as a number $B^{r_A\rightarrow 1^{\vec{n}_A}}$. It is evident that
	\begin{equation}
		B^{a\rightarrow 1^{(2,0)}}=B^{b\rightarrow 1^{(0,2)}}=1
	\end{equation}
	because the anti-trivial representation $1^{(2,0)}$ (or $1^{(0,2)}$) of $S_2$ in $a$ (or $b$) is just itself. In addition, we have 
	\begin{equation}
		B^{d\rightarrow 1^{(1,1)}}=B^{e\rightarrow 1^{(1,1)}}=\pm 1
	\end{equation}
	because $\ket{d\rightarrow 1^{(1,1)}}=\pm\ket{d}$ (or $\ket{e\rightarrow 1^{(1,1)}}=\pm\ket{e}$) can furnish the anti-trivial representation $1^{(1,1)}$ of $S_{1}\times S_{1}$ in $d$ (or $e$). This arbitrariness will not cause any problem as the branching coefficient always appears twice in the group theoretical coefficient. The above four branching coefficients are all we need. Now, the relevant group theoretical coefficients are given by
	\begin{equation}
		\begin{aligned}
			&C^{a}(1)=|S_2| \sqrt{\frac{d_{a}}{2!}}\text{sgn}(1)\Gamma^{a}(1)B^{a\rightarrow 1^{(2,0)}} B^{a\rightarrow 1^{(2,0)}}\\
			&C^{b}(1)=|S_2| \sqrt{\frac{d_{b}}{2!}}\text{sgn}(1)\Gamma^{b}(1)B^{b\rightarrow 1^{(0,2)}} B^{b\rightarrow 1^{(0,2)}}\\
			&C^{d}(1)=|S_1\times S_1| \sqrt{\frac{d_{d}}{2!}}\text{sgn}(1)\Gamma^{d}(1)B^{d\rightarrow 1^{(1,1)}} B^{d\rightarrow 1^{(1,1)}}\\
			&C^{d}((12))=|S_1\times S_1| \sqrt{\frac{d_{d}}{2!}}\text{sgn}((12))\Gamma^{d}((12))B^{d\rightarrow 1^{(1,1)}} B^{d\rightarrow 1^{(1,1)}}\\
			&C^{e}(1)=|S_1\times S_1| \sqrt{\frac{d_{e}}{2!}}\text{sgn}(1)\Gamma^{e}(1)B^{e\rightarrow 1^{(1,1)}} B^{e\rightarrow 1^{(1,1)}}\\
			&C^{e}((12))=|S_1\times S_1| \sqrt{\frac{d_{e}}{2!}}\text{sgn}((12))\Gamma^{e}((12))B^{e\rightarrow 1^{(1,1)}} B^{e\rightarrow 1^{(1,1)}}\\
		\end{aligned}
	\end{equation}
	where the elements of $S_2$ are denoted by $1$ (the identity) as well as $(12)$, and we use the fact that $\Gamma^{r^T}(\sigma)=\text{sgn}(\sigma)\Gamma^{r}(\sigma)$ for the symmetric group. The final results are
	\begin{equation}
		\begin{aligned}
			&C^{a}(1)=\sqrt{2}; \quad C^{b}(1)=\sqrt{2}; \\
			&C^{d}(1)=\frac{1}{\sqrt{2}}; \quad C^{d}((12))=-\frac{1}{\sqrt{2}}\\
			&C^{e}(1)=\frac{1}{\sqrt{2}}; \quad C^{e}((12))=\frac{1}{\sqrt{2}}
		\end{aligned}
	\end{equation}
	Notice that $C^{a}((12))$ and $C^{b}((12))$ are not needed. To understand why this is the case, recall that the LHS of the basis transformation shown in equation (\ref{basis transformation}), requires only Gauss graph operators with non-equivalent $\sigma_A$. With $r_A=a$, we have $H_{\vec{n}_A}=S_2$ so that all elements of $S_2$ are in the same equivalence class of the double coset $S_2\backslash S_2/ S_2$. Thus for $r_A=a$, we only need operators with $\sigma_A=1$. In fact, a simple calculation shows that $C^{a}(1)=C^{a}((12))$ as we expect. The same argument shows that $C^{b}((12))$ is not needed for the same reason. However, for $r_A=d$ or $e$, we have $H_{\vec{n}_A}=S_1\times S_1$. In this case, 1 and (12) are in different classes of the double coset $(S_1\times S_1)\backslash S_2 / (S_1\times S_1)$ so that they are not equivalent. Hence, we need both operators with $\sigma_A=1$ as well as $\sigma_A=(12)$.
	
	Finally, as we will use Gauss graph operators to diagonalize the dilatation operator, we implicitly assume that $D_{31}$ and $D_{21}$ commute with $D_{32}$. And we find this is true by using the exact result we obtain to calculate the commutators---matrix elements in commutators $[D_{31},D_{32}]$ and $[D_{21},D_{32}]$ are of order higher or equal to $1/(\xi_0\xi_1^2)$ in the expansion introduced above. Thus, they vanish at the order we consider.
	
	These details give everything that is needed to carry out the calculation We give the results in the next subsection.
	
	\subsection{Results at Leading Order}
	
	To express the result for the action of the dilatation operator in the Gauss graph basis, it is worth introducing a concise notation to specify the relevant Gauss graph operators. We use $\alpha,\beta,\gamma$ to specify $\vec{n}_A$ where
	\begin{equation}
		\alpha=(2,0); \quad \beta=(0,2); \quad \gamma=(1,1)
	\end{equation}
	For example, $O_{(\xi_0,\xi_1)\gamma\alpha}((12),1)$ specifies the Gauss graph operator with $\vec{n}_2=\gamma,\vec{n}_3=\alpha,\sigma_{2}=(12),\sigma_{3}=1$ and $r_1=(\xi_0,\xi_1)$. With the help of equation (\ref{basis transformation}), we see that this Gauss graph operator is given by
	\begin{equation}
		\begin{aligned}
			O_{(\xi_0,\xi_1)\gamma\alpha}((12),1)&=C^{d}((12))C^{a}(1)O_{da}(\xi_0,\xi_1)+C^{e}((12))C^{a}(1)O_{ea}(\xi_0,\xi_1)\\
			&=-O_{da}(\xi_0,\xi_1)+O_{ea}(\xi_0,\xi_1)
		\end{aligned}
	\end{equation}
	It is evident that we have 16 Gauss graph operators. In the notation we have just introduced, they can be written as
	\begin{equation}
		\begin{array}{llll}
			O_{(\xi_0,\xi_1)\alpha\alpha}(1,1); & O_{(\xi_0,\xi_1)\gamma\alpha}(1,1); & O_{(\xi_0,\xi_1)\gamma\alpha}((12),1); & O_{(\xi_0,\xi_1)\beta\alpha}(1,1);\\
			O_{(\xi_0,\xi_1)\alpha\gamma}(1,1); & O_{(\xi_0,\xi_1)\alpha\gamma}(1,(12)); & O_{(\xi_0,\xi_1)\gamma\gamma}(1,1); & O_{(\xi_0,\xi_1)\gamma\gamma}(1,(12));\\
			O_{(\xi_0,\xi_1)\gamma\gamma}((12),1); & O_{(\xi_0,\xi_1)\gamma\gamma}((12),(12)); & O_{(\xi_0,\xi_1)\beta\gamma}(1,1); & O_{(\xi_0,\xi_1)\beta\gamma}(1,(12));\\
			O_{(\xi_0,\xi_1)\alpha\beta}(1,1); & O_{(\xi_0,\xi_1)\gamma\beta}(1,1); & O_{(\xi_0,\xi_1)\gamma\beta}((12),1); & O_{(\xi_0,\xi_1)\beta\beta}(1,1);\\
		\end{array}
	\end{equation}
	Each of these operators is specified by a Gauss graph. We will soon spell this correspondence out in detail. In terms of this notation, the action of the dilatation operator in the Gauss graph basis calculated analytically is shown below. We have considered the large $N$ limit and the displaced corners limit, hence we show the leading order result. Only the non-vanishing matrix elements are shown. For example, our analytic result implies that $D_{31}O_{(\xi_0,\xi_1)\gamma\alpha}(1,1)=0$, so we will not write it down. Also, to simplify our expressions we use shorthand $C_0=N-\xi_0$ and $C_1=N-\xi_0-\xi_1$.
	
	The result of the action of $D_{31}$ in the Gauss graph basis reads
	\begin{equation}
		\begin{aligned}
			&D_{31} O_{(\xi_0,\xi_1)\alpha\gamma}(1,(12))=2\left[  -(C_0+C_1)O_{(\xi_0,\xi_1)\alpha\gamma}(1,(12))  \right.\\
			&\left.+\sqrt{C_0C_1}O_{(\xi_0+1,\xi_1-2)\alpha\gamma}(1,(12))+\sqrt{C_0C_1}O_{(\xi_0-1,\xi_1+2)\alpha\gamma}(1,(12))\right]
		\end{aligned}
	\end{equation}
	\begin{equation}
		\begin{aligned}
			&D_{31} O_{(\xi_0,\xi_1)\gamma\gamma}(1,(12))=2\left[  -(C_0+C_1)O_{(\xi_0,\xi_1)\gamma\gamma}(1,(12))  \right.\\
			&\left.+\sqrt{C_0C_1}O_{(\xi_0+1,\xi_1-2)\gamma\gamma}(1,(12))+\sqrt{C_0C_1}O_{(\xi_0-1,\xi_1+2)\gamma\gamma}(1,(12))\right]
		\end{aligned}
	\end{equation}
	\begin{equation}
		\begin{aligned}
			&D_{31} O_{(\xi_0,\xi_1)\gamma\gamma}((12),(12))=2\left[  -(C_0+C_1)O_{(\xi_0,\xi_1)\gamma\gamma}((12),(12))  \right.\\
			&\left.+\sqrt{C_0C_1}O_{(\xi_0+1,\xi_1-2)\gamma\gamma}((12),(12))+\sqrt{C_0C_1}O_{(\xi_0-1,\xi_1+2)\gamma\gamma}((12),(12))\right]
		\end{aligned}
	\end{equation}
	\begin{equation}
		\begin{aligned}
			&D_{31} O_{(\xi_0,\xi_1)\beta\gamma}(1,(12))=2\left[  -(C_0+C_1)O_{(\xi_0,\xi_1)\beta\gamma}(1,(12))  \right.\\
			&\left.+\sqrt{C_0C_1}O_{(\xi_0+1,\xi_1-2)\beta\gamma}(1,(12))+\sqrt{C_0C_1}O_{(\xi_0-1,\xi_1+2)\beta\gamma}(1,(12))\right]
		\end{aligned}
	\end{equation}
	
	The result of the action of $D_{21}$ in the Gauss graph basis reads 
	\begin{equation}
		\begin{aligned}
			&D_{21} O_{(\xi_0,\xi_1)\gamma\alpha}((12),1)=2\left[  -(C_0+C_1)O_{(\xi_0,\xi_1)\gamma\alpha}((12),1)  \right.\\
			&\left.+\sqrt{C_0C_1}O_{(\xi_0+1,\xi_1-2)\gamma\alpha}((12),1)+\sqrt{C_0C_1}O_{(\xi_0-1,\xi_1+2)\gamma\alpha}((12),1)\right]
		\end{aligned}
	\end{equation}
	\begin{equation}
		\begin{aligned}
			&D_{21} O_{(\xi_0,\xi_1)\gamma\gamma}((12),1)=2\left[  -(C_0+C_1)O_{(\xi_0,\xi_1)\gamma\gamma}((12),1)  \right.\\
			&\left.+\sqrt{C_0C_1}O_{(\xi_0+1,\xi_1-2)\gamma\gamma}((12),1)+\sqrt{C_0C_1}O_{(\xi_0-1,\xi_1+2)\gamma\gamma}((12),1)\right]
		\end{aligned}
	\end{equation}
	\begin{equation}
		\begin{aligned}
			&D_{21} O_{(\xi_0,\xi_1)\gamma\gamma}((12),(12))=2\left[  -(C_0+C_1)O_{(\xi_0,\xi_1)\gamma\gamma}((12),(12))  \right.\\
			&\left.+\sqrt{C_0C_1}O_{(\xi_0+1,\xi_1-2)\gamma\gamma}((12),(12))+\sqrt{C_0C_1}O_{(\xi_0-1,\xi_1+2)\gamma\gamma}((12),(12))\right]
		\end{aligned}
	\end{equation}
	\begin{equation}
		\begin{aligned}
			&D_{21} O_{(\xi_0,\xi_1)\gamma\beta}((12),1)=2\left[  -(C_0+C_1)O_{(\xi_0,\xi_1)\gamma\beta}((12),1)  \right.\\
			&\left.+\sqrt{C_0C_1}O_{(\xi_0+1,\xi_1-2)\gamma\beta}((12),1)+\sqrt{C_0C_1}O_{(\xi_0-1,\xi_1+2)\gamma\beta}((12),1)\right]
		\end{aligned}
	\end{equation}
	
	The result of the action of $D_{32}$ in the Gauss graph basis reads 
	\begin{equation}\label{comparision eq1}
		\begin{aligned}
			D_{32}O_{(\xi_0,\xi_1)\gamma\alpha}((12),1)&=\frac{4C_1}{\xi_0}O_{(\xi_0,\xi_1)\gamma\alpha}((12),1)\\
			&-\frac{2\sqrt{2}\sqrt{C_0C_1}}{\xi_0}O_{(\xi_0,\xi_1)\gamma\gamma}((12),1)\\
		\end{aligned}
	\end{equation}
	\begin{equation}
		\begin{aligned}
			D_{32}O_{(\xi_0,\xi_1)\alpha\gamma}(1,(12))&=\frac{4C_1}{\xi_0}O_{(\xi_0,\xi_1)\alpha\gamma}(1,(12))\\
			&-\frac{2\sqrt{2}\sqrt{C_0C_1}}{\xi_0}O_{(\xi_0,\xi_1)\gamma\gamma}(1,(12))\\
		\end{aligned}
	\end{equation}
	\begin{equation}
		\begin{aligned}
			D_{32}O_{(\xi_0,\xi_1)\gamma\gamma}(1,(12))&=\frac{2(C_0+C_1)}{\xi_0}O_{(\xi_0,\xi_1)\gamma\gamma}(1,(12))\\
			&-\frac{2\sqrt{2}\sqrt{C_0C_1}}{\xi_0}O_{(\xi_0,\xi_1)\alpha\gamma}(1,(12))\\
			&-\frac{2\sqrt{2}\sqrt{C_0C_1}}{\xi_0}O_{(\xi_0,\xi_1)\beta\gamma}(1,(12))\\
		\end{aligned}
	\end{equation}
	\begin{equation}
		\begin{aligned}
			D_{32}O_{(\xi_0,\xi_1)\gamma\gamma}((12),1)&=\frac{2(C_0+C_1)}{\xi_0}O_{(\xi_0,\xi_1)\gamma\gamma}((12),1)\\
			&-\frac{2\sqrt{2}\sqrt{C_0C_1}}{\xi_0}O_{(\xi_0,\xi_1)\gamma\alpha}((12),1)\\
			&-\frac{2\sqrt{2}\sqrt{C_0C_1}}{\xi_0}O_{(\xi_0,\xi_1)\gamma\beta}((12),1)\\
		\end{aligned}
	\end{equation}
	\begin{equation}
		\begin{aligned}
			D_{32}O_{(\xi_0,\xi_1)\gamma\gamma}((12),(12))&=\frac{2(C_0+C_1)}{\xi_0}O_{(\xi_0,\xi_1)\gamma\gamma}((12),(12))\\
		\end{aligned}
	\end{equation}
	\begin{equation}
		\begin{aligned}
			D_{32}O_{(\xi_0,\xi_1)\beta\gamma}(1,(12))&=\frac{4C_0}{\xi_0}O_{(\xi_0,\xi_1)\beta\gamma}(1,(12))\\
			&-\frac{2\sqrt{2}\sqrt{C_0C_1}}{\xi_0}O_{(\xi_0,\xi_1)\gamma\gamma}(1,(12))\\
		\end{aligned}
	\end{equation}
	\begin{equation}
		\begin{aligned}
			D_{32}O_{(\xi_0,\xi_1)\gamma\beta}((12),1)&=\frac{4C_0}{\xi_0}O_{(\xi_0,\xi_1)\gamma\beta}((12),1)\\
			&-\frac{2\sqrt{2}\sqrt{C_0C_1}}{\xi_0}O_{(\xi_0,\xi_1)\gamma\gamma}((12),1)\\
		\end{aligned}
	\end{equation}
	
	Each of the above operators is normalized. This result is in perfect agreement with the analytic formulas given in equations (\ref{D31,D21 in Gauss graph basis}) and $(\ref{M32 in Gauss graph basis})$, providing a highly non-trivial check of our computation. We are now ready to extract the subleading corrections to the above results.
	
	\subsection{The Subleading Interaction}\label{sbldin}
	
	Our main interest in this section, is the subleading contribution to the action of $D_{32}$, which describes interactions of the excitations of branes, given by the edges of the Gauss graphs. The leading order result is expected to have the form given in equation (\ref{M32 in Gauss graph basis}), which has been argued to arise from the interaction Hamiltonian $\text{Tr}\left([\bar{b}, \bar{a}][a, b]\right)$. Our analytical calculation has verified this expectation. 
	
	We will now relax the strict displaced corners limit and evaluate the first corrections which appear. Of course, by moving further and further from the displaced corners limit we will reach a point where our formulas break down and the Gauss graph operators are no longer well defined. However close to the displaced corners limit we expect our description remains sensible, basically because we have confirmed the duality between our dilatation operator and a system of giant graviton branes, owning a nice semi-classical description in terms of branes excited by open strings.
	
	We now proceed to calculate the subleading correction to $D_{32}$. Our two giant graviton system again provides the simplest possible setting for this task. The subleading correction is given by expanding the matrix elements of $D_{32}$ in terms of $\frac{1}{\xi_1}$ around $\xi_1=\infty$ and retaining terms of order $\mathcal{O}(\frac{1}{\xi_1})$. We will assume that we are in the strict large $N$ limit, so that we set $C_1\rightarrow C_0$. This expansion yields
	\begin{equation}
		D_{32}=D_{32}^{(0)}+\frac{C_0}{\xi_0 \xi_1}D_{32}^{(1)}
	\end{equation}
	The action of $D_{32}^{(0)}$ is the leading action of the dilatation operator, given in the previous section.
	The subleading correction we obtain is shown in what follows

	\begin{equation}
		\begin{aligned}
			&D_{32}^{(1)}O_{(\xi_0,\xi_1)\alpha\alpha}(1,1)=0
		\end{aligned}
	\end{equation}
	{\vskip 0.3cm}
	
	\begin{equation}
		\begin{aligned}
			D_{32}^{(1)}O_{(\xi_0,\xi_1)\gamma\alpha}(1,1)&=-4O_{(\xi_0,\xi_1)\alpha\gamma}(1,(12))\\
			& +2\sqrt{2}O_{(\xi_0,\xi_1)\gamma\gamma}(1,(12))\\
		\end{aligned}
	\end{equation}
	{\vskip 0.3cm}
	
	\begin{equation}\label{Ogammaalpha121}
		\begin{aligned}
			&D_{32}^{(1)}O_{(\xi_0,\xi_1)\gamma\alpha}((12),1)=-4O_{(\xi_0,\xi_1)\gamma\alpha}((12),1)\\
		\end{aligned}
	\end{equation}
	{\vskip 0.3cm}
	
	\begin{equation}
		\begin{aligned}
			D_{32}^{(1)}O_{(\xi_0,\xi_1)\beta\alpha}(1,1)&=-4O_{(\xi_0,\xi_1)\gamma\gamma}(1,(12))\\
			&+4\sqrt{2}O_{(\xi_0,\xi_1)\beta\gamma}(1,(12))
		\end{aligned}
	\end{equation}
	{\vskip 0.3cm}
	
	\begin{equation}
		\begin{aligned}
			D_{32}^{(1)}O_{(\xi_0,\xi_1)\alpha\gamma}(1,1)&=4O_{(\xi_0,\xi_1)\alpha\gamma}(1,(12))\\
			&-2\sqrt{2}O_{(\xi_0,\xi_1)\gamma\gamma}(1,(12))
		\end{aligned}
	\end{equation}
	{\vskip 0.3cm}
	
	\begin{equation}
		\begin{aligned}
			D_{32}^{(1)}O_{(\xi_0,\xi_1)\alpha\gamma}(1,(12))&=-4O_{(\xi_0,\xi_1)\gamma\alpha}(1,1) \\
			&+4O_{(\xi_0,\xi_1)\alpha\gamma}(1,1)\\
			&+4O_{(\xi_0,\xi_1)\alpha\gamma}(1,(12)) \\
			&+2\sqrt{2}O_{(\xi_0,\xi_1)\gamma\gamma}(1,1)\\
			&+2\sqrt{2}O_{(\xi_0,\xi_1)\gamma\gamma}((12),(12))\\
			&-4\sqrt{2}O_{(\xi_0,\xi_1)\alpha\beta}(1,1)\\
		\end{aligned}
	\end{equation}
	{\vskip 0.3cm}
	
	\begin{equation}
		\begin{aligned}
			D_{32}^{(1)}O_{(\xi_0,\xi_1)\gamma\gamma}(1,1)&=2\sqrt{2}O_{(\xi_0,\xi_1)\alpha\gamma}(1,(12))\\
			&-2\sqrt{2}O_{(\xi_0,\xi_1)\beta\gamma}(1,(12))\\
		\end{aligned}
	\end{equation}
	{\vskip 0.3cm}
	
	\begin{equation}
		\begin{aligned}
			D_{32}^{(1)}O_{(\xi_0,\xi_1)\gamma\gamma}(1,(12))&=2\sqrt{2}O_{(\xi_0,\xi_1)\gamma\alpha}(1,1)\\
			&-2\sqrt{2}O_{(\xi_0,\xi_1)\alpha\gamma}(1,1)\\
			&-4O_{(\xi_0,\xi_1)\beta\alpha}(1,1)\\
			&+4O_{(\xi_0,\xi_1)\alpha\beta}(1,1)\\
			&+2\sqrt{2}O_{(\xi_0,\xi_1)\beta\gamma}(1,1)\\
			&-2\sqrt{2}O_{(\xi_0,\xi_1)\gamma\beta}(1,1)\\
		\end{aligned}
	\end{equation}
	{\vskip 0.3cm}
	
	\begin{equation}
		\begin{aligned}
			D_{32}^{(1)}O_{(\xi_0,\xi_1)\gamma\gamma}((12),1)&=0
		\end{aligned}
	\end{equation}
	{\vskip 0.3cm}
	
	\begin{equation}
		\begin{aligned}
			D_{32}^{(1)}O_{(\xi_0,\xi_1)\gamma\gamma}((12),(12))&=2\sqrt{2}O_{(\xi_0,\xi_1)\alpha\gamma}(1,(12))\\
			&-2\sqrt{2}O_{(\xi_0,\xi_1)\beta\gamma}(1,(12))\\
		\end{aligned}
	\end{equation}
	{\vskip 0.3cm}
	
	\begin{equation}
		\begin{aligned}
			D_{32}^{(1)}O_{(\xi_0,\xi_1)\beta\gamma}(1,1)&=2\sqrt{2}O_{(\xi_0,\xi_1)\gamma\gamma}(1,(12))\\
			&-4O_{(\xi_0,\xi_1)\beta\gamma}(1,(12))
		\end{aligned}
	\end{equation}
	{\vskip 0.3cm}
	
	\begin{equation}
		\begin{aligned}
			D_{32}^{(1)}O_{(\xi_0,\xi_1)\beta\gamma}(1,(12))&=4\sqrt{2}O_{(\xi_0,\xi_1)\beta\alpha}(1,1)\\
			&-2\sqrt{2}O_{(\xi_0,\xi_1)\gamma\gamma}(1,1)\\
			&-2\sqrt{2}O_{(\xi_0,\xi_1)\gamma\gamma}((12),(12))\\ 
			&-4O_{(\xi_0,\xi_1)\beta\gamma}(1,1)\\
			&-4O_{(\xi_0,\xi_1)\beta\gamma}(1,(12))\\
			&+4O_{(\xi_0,\xi_1)\gamma\beta}(1,1)\\
		\end{aligned}
	\end{equation}
	{\vskip 0.3cm}
	
	\begin{equation}
		\begin{aligned}
			D_{32}^{(1)}O_{(\xi_0,\xi_1)\alpha\beta}(1,1)&=-4\sqrt{2}O_{(\xi_0,\xi_1)\alpha\gamma}(1,(12))\\
			&+4O_{(\xi_0,\xi_1)\gamma\gamma}(1,(12))
		\end{aligned}
	\end{equation}
	{\vskip 0.3cm}
	
	\begin{equation}
		\begin{aligned}
			D_{32}^{(1)}O_{(\xi_0,\xi_1)\gamma\beta}(1,1)&=-2\sqrt{2}O_{(\xi_0,\xi_1)\gamma\gamma}(1,(12))\\
			&+4O_{(\xi_0,\xi_1)\beta\gamma}(1,(12))\\
		\end{aligned}
	\end{equation}
	{\vskip 0.3cm}
	
	\begin{equation}
		\begin{aligned}
			D_{32}^{(1)}O_{(\xi_0,\xi_1)\gamma\beta}((12),1)&=4O_{(\xi_0,\xi_1)\gamma\beta}((12),1)\\
		\end{aligned}
	\end{equation}
	{\vskip 0.3cm}
	
	\begin{equation}
		\begin{aligned}
			D_{32}^{(1)}O_{(\xi_0,\xi_1)\beta\beta}(1,1)&=0\\
		\end{aligned}
	\end{equation}
	{\vskip 0.3cm}
	
	The fact that the matrix of this subleading correction is symmetric implies that all corrections to the anomalous dimensions are real, as they must be. To interpret this result, it is necessary to write it in terms of the oscillators of the emergent gauge theory. Remarkably, this can be done and one finds the following interaction term written in terms of oscillators reproduces the matrix elements of the subleading action of $D_{32}$ 
	\begin{equation}
		\begin{aligned}
			D_{32}^{(1)}=2{\rm Tr}&\Big(\sigma_z \left(\bar{b}_2 b_2 \bar{b}_3 b_3 \bar{b}_3 b_3 + \bar{b}_3 \bar{b}_2 b_2 \bar{b}_3 b_3 b_3 + \bar{b}_3 \bar{b}_3 b_3 \bar{b}_2 b_2 b_3-\bar{b}_2 \bar{b}_3 b_2 \bar{b}_3 b_3 b_3  \right.\\
			&\left.  -\bar{b}_3 \bar{b}_2 b_3 \bar{b}_3 b_2 b_3 - \bar{b}_3 \bar{b}_3 b_3 \bar{b}_2 b_3 b_2 +\bar{b}_2 \bar{b}_2 \bar{b}_3 b_3 b_2 b_2 - \bar{b}_2 \bar{b}_2 b_2 b_2 \hat{n}_3 \right)\Big)\\
			+2{\rm Tr}&\Big(\sigma_x\sigma_z\left(\hat{n}_2\bar{b}_3\bar{b}_3b_3b_3-\bar{b}_3\bar{b}_3b_3b_3\hat{n}_2 +\bar{b}_3\bar{b}_3b_3\hat{n}_2b_3 -\bar{b}_3\hat{n}_2\bar{b}_3b_3b_3 \right)\Big)\\
			-4{\rm Tr}&\Big(\sigma_x\hat{n}_2\sigma_z\hat{n}_3\Big)
		\end{aligned}\label{subint}
	\end{equation}
	where we have used the usual Pauli matrices 
	$$
	\sigma_z=\left[\begin{matrix} 1 &0\\ 0 &-1\end{matrix}\right]\qquad
	\sigma_x=\left[\begin{matrix} 0 &1\\ 1 &0\end{matrix}\right]
	$$
	By identifying $\bar{b},b$ with $\bar{b}_3,b_3$ and $\bar{a},a$ with $\bar{b}_2,b_2$ as we did in the previous section, the above formula gives the interactions of the emergent Yang-Mills theory.
	
	The emergent Yang-Mills theory lives on the world volume of two giant graviton branes. As a result, the gauge group is U(2).
	These branes have different radii implying that the open strings stretching between the branes are massive. The strings tretching from a giant graviton brane, back to the same brane remain massless. Thus, the emergent guage theory is on the Coulomb branch and the U(2) gauge symmetry is broken to U(1)$\times$U(1). Under this gauge group we find
	\begin{equation}
		b_2\to e^{i\theta_1}b_2\qquad b_3\to e^{i\theta_2}b_3 \qquad \hat{n}_2\to\hat{n}_2\qquad \hat{n}_3\to\hat{n}_3
	\end{equation}
	
	It is not hard to check that (\ref{subint}) is indeed invariant under these U(1)$\times$U(1) transformations.
	
	At leading order, the mixing is between $\phi_1$ and $\phi_2$, as well as between $\phi_1$ and $\phi_3$. This mixing is diagonalized by the Gauss graph operators. The leading contribution to the mixing involving $\phi_2$ and $\phi_3$ weakly mixes operators labeled by distinct Gauss graphs: two operators can only mix if they differ, at most, by the placement of a single edge that has both end points attached to a single node. The interaction in (\ref{subint}) allows graphs to mix even if this involves rearranging open edges. In Figure 1 below we have shown four pairs of graphs that are mixed by (\ref{subint}), but are not mixed by the leading contribution to the mixing involving $\phi_2$ and $\phi_3$. The interaction allows both the movement of closed loop edges from one node to another as well as the rearrangement of closed loop edges with both ends at the same node, into open edges that have their endpoints at different nodes.
	
	\begin{figure}[h]
		\centering
		\includegraphics[width=1.0\linewidth]{"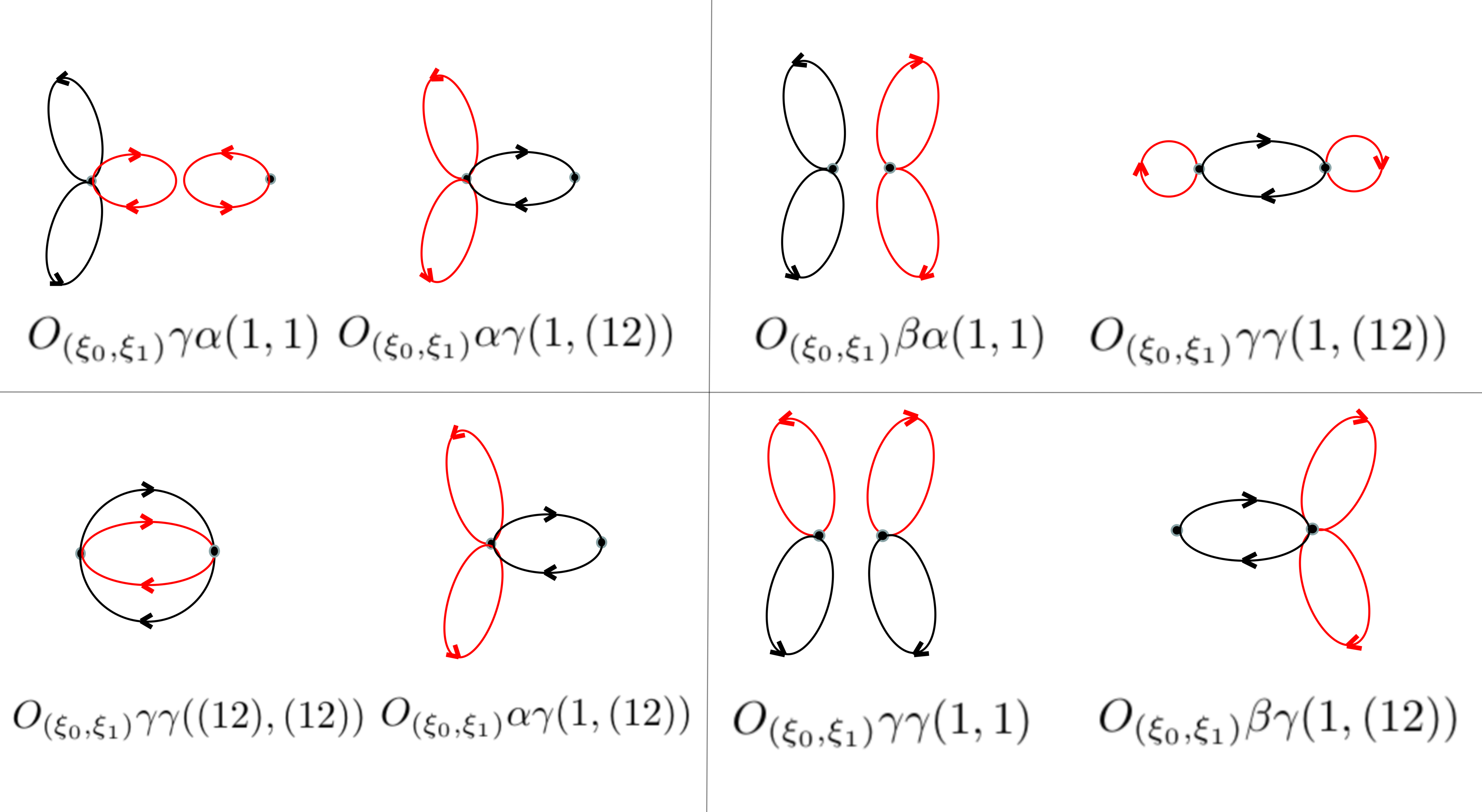"}
		\caption{Each pair of graphs shown are mixed by (\ref{subint}). They are not mixed by the leading contribution to the mixing involving $\phi_2$ and $\phi_3$. The edges for $\phi_{2}$ are colored red while edges for $\phi_{3}$ are colored black.}
		\label{fig:gauss-graphs}
	\end{figure}

	\subsection{Numerical Spectra}
	
	In this subsection, we discuss the eigenvalues of $D_{32}$, which is the Hamiltonian describing the interactions of the excitations. Let $n_T$ denote the total number of boxes in $R$, while $N$ is rank of the gauge group as usual. There are bounds for $n_T$ as follows 
	\begin{equation}
		N<n_T<2N
	\end{equation}
	The lower limit sets the smallest radius possible for the giant gravitons two giant graviton system. This corresponds to the case that each column has $N/2$ boxes, so that both still have a macroscopic size. The upper limit reflects the restriction that the number of rows must be less than $N$ and the Young diagrams we consider have only two columns. 
	These bounds also agree with the fact that we are considering operators with a $\sim N$ dimension so that the Young diagrams labeling them have $\sim N$ boxes.
	
\begin{figure}[h]
		\centering
		\includegraphics[width=1.0\linewidth]{"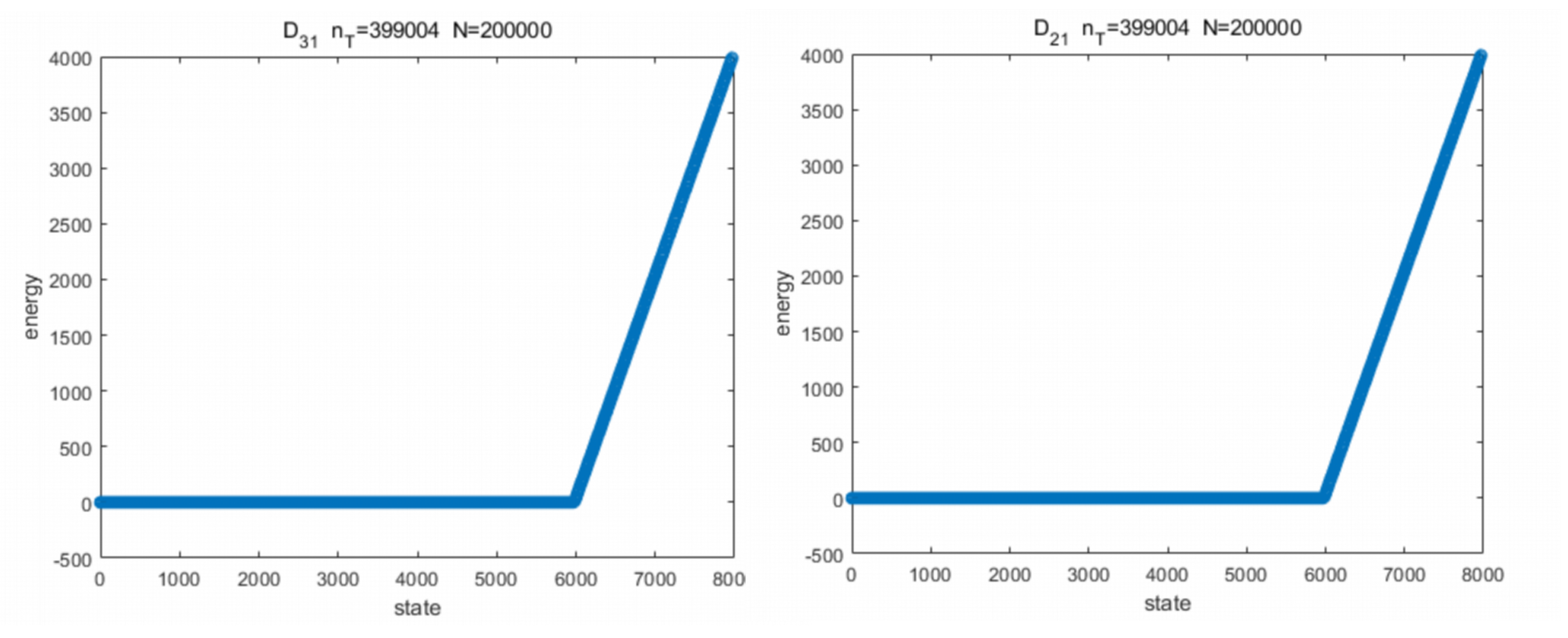"}
		\caption{The first plot is the spectrum of leading action of $D_{31}$, while the second plot is that of $D_{21}$. Each spectrum is calculated in the system with $n_T$=399004 and $N=200000$.}
		\label{fig:leading-d31-d21-nt399004-n200000-v1}
	\end{figure}

	It is easy to verify that $D_{32}$, $D_{31}$ and $D_{21}$ commutes, so that it is sensible to consider the spectrum from a given one of them. Before considering the spectra of $D_{32}$, consider the spectra of $D_{31}$ and $D_{21}$. Examples of spectra are given in Fig \ref{fig:leading-d31-d21-nt399004-n200000-v1}. This shows that the system is a harmonic oscillator with a high energy cut off, in perfect agreement with the results of \cite{2010Emergentthree}.
	
\begin{figure}[h]
		\centering
		\includegraphics[width=1.0\linewidth]{"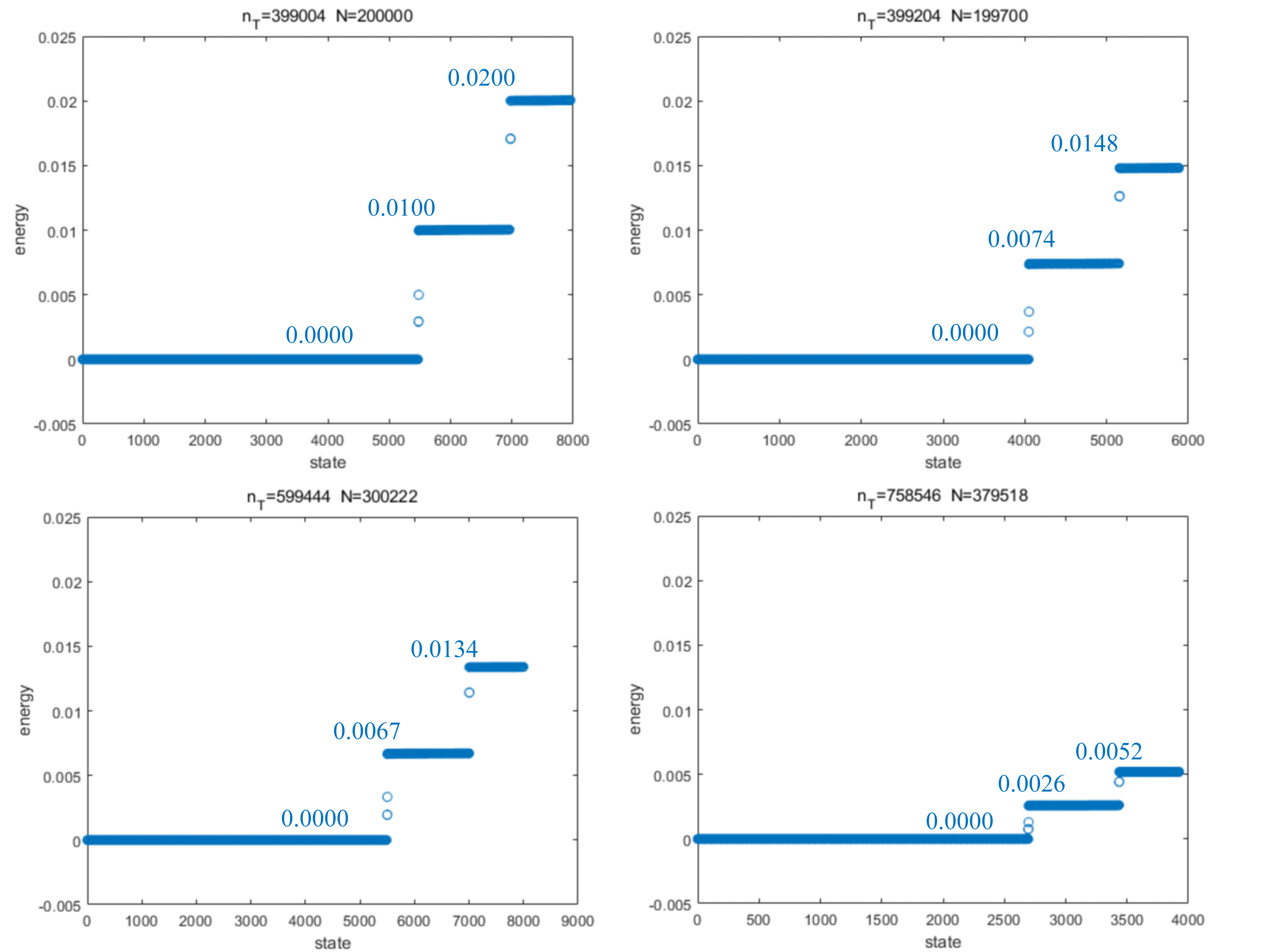"}
		\caption{The parameters $n_T$ and $N$ of the system are shown above each spectrum. It is easy to verify that the intervals between adjacent energy levels are constant in each spectrum.}
		\label{fig:leading-spectrum01}
	\end{figure}

	Now consider the spectrum of the leading contribution to $D_{32}$. Again the spectrum is very similar to that of the oscillator, i.e. the spectrum has several evenly spaced energy levels. The fact that there are only two levels is also not surprising and it reflects the fact that each operator is constructed using only 2 $\phi_2$ fields and 2 $\phi_3$ fields. See Figure \ref{fig:leading-spectrum01} for some examples of typical spectra. This sets the contribution to the energy levels of the system. 
	
	\begin{table}[h]
		\centering
		\begin{tabular}{|l|l|l|l|}
			\hline
			$(N,n_T)$& $D_0$  &  $D_1$  & $D_2$ \\ \hline
			(200000,399004)& 5476 & 1495 & 994 \\ \hline
			(199700,399204)& 4046 & 1105 & 734 \\ \hline
			(300222,599444)& 5498 & 1501 & 998 \\ \hline
			(379518,758546)& 2696 & 736 & 488 \\ \hline
		\end{tabular}
		\caption{The table presents the degeneracies of energy levels of each spectrum shown in Figure \ref{fig:leading-spectrum01}. We show pairs of parameters $N$ and $n_T$ used to compute each spectrum in the first column, while degeneracies of corresponding energy levels are shown in the latter three columns. Notation $D_0, D_1, D_2$ respectively denote the degeneracy of energy level $0, E_{0}, 2E_{0}$.}
		\label{table: degeneracies}
	\end{table}

	The complete spectrum takes the form $0,E^{(0)},2E^{(0)}$ with the size of $E^{(0)}$ set by the formula ($\xi_{1,\text{max}}=2N-n_T$)
	\begin{equation}\label{E0 of N and nT}
		E^{(0)}=2\left( \frac{N}{n_T-N}-1 \right) =2\frac{\xi_{1,\text{max}}}{N-\xi_{1,\text{max}}}
	\end{equation}
	Notice that $E^{(0)}$ is an order 1 number as we take $N\to\infty$. Some numerical results of $E^{(0)}$ are shown in Table \ref{table: E0}. A simple check shows that formula \eqref{E0 of N and nT} is in a perfect agreement with these numerical results, within an accuracy of $1\times10^{-4}$. To facilitate the study on the spectrum of the system, in Table \ref{table: degeneracies} we also present the degeneracies of energy levels shown in Figure \ref{fig:leading-spectrum01}. All degeneracies listed in the table are exact to the precision we use, except for the second level $2E^{(0)}=0.0200$ of the spectrum computed with parameters $N=200000,n_T=399004$. We find there are 796 states of energy 0.0200 and 198 states of energy 0.0201. But, as such deviation is rather small, we still assume the second level of this spectrum is 0.0200. In fact, the average energy of these 796+198 states is 0.0200, as we expect.

\begin{figure}[h]
		\centering
		\includegraphics[width=1.0\linewidth]{"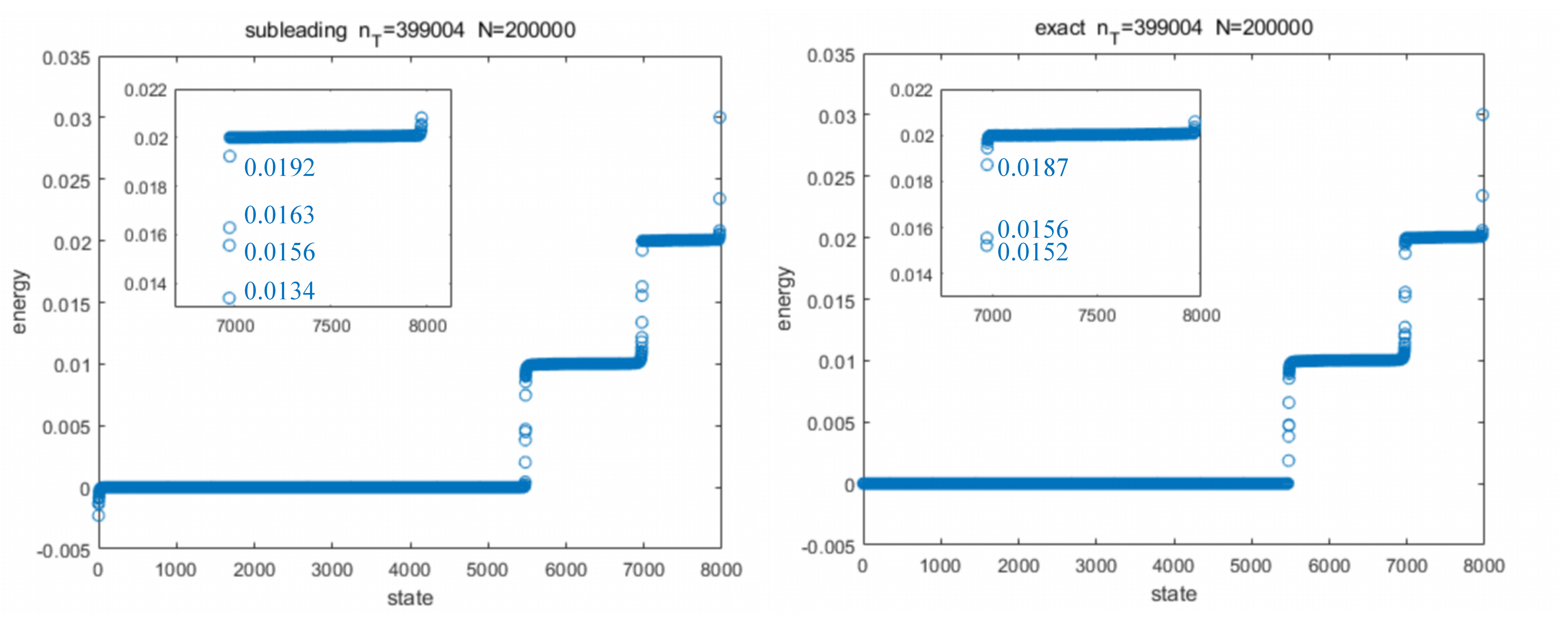"}
		\caption{The spectra are calculated in the system with $n_T=399004$ and $N=200000$. The first plot is the spectrum of action with subleading corrections, while the second one is that of exact action. New energies caused by subleading corrections are 0.0134 and 0.0192, as 0.0156 and 0.0163 can also be found in the spectrum of leading action shown in Fig \ref{fig:leading-spectrum01}, including acceptable errors.}
		\label{fig:subleading-exact-spectrum-nt399004-n200000011}
	\end{figure}

	Now consider the contribution to the spectrum coming from subleading corrections. An example is given in Fig \ref{fig:subleading-exact-spectrum-nt399004-n200000011}. Comparing it with the leading spectrum shown in \ref{fig:leading-spectrum01}, we see that these subleading corrections give a very small correction to the leading energy level. Simply based on this numerical evidence, we confirm that our expansion converges very rapidly. Again, in Table \ref{table: subleading degeneracies} we show degeneracies of dominating energy levels in Figure \ref{fig:subleading-exact-spectrum-nt399004-n200000011}. There are also a relatively small number of states of energy 0.0201 in each spectrum. But the average energy is still 0.0200 when those states of energy 0.0200 are included. 
		
	\begin{table}[h]
		\centering
		\begin{tabular}{|l|l|l|l|l|l|l|}
			\cline{1-3} \cline{5-7}
			$N$                            & $n_T$   & $E^{(0)}$      &  & $N$      & $n_T$   & $E^{(0)}$      \\ \cline{1-3} \cline{5-7} 
			200100                       & 399004 & 0.0120 &  & 300100 & 599004 & 0.0080 \\ \cline{1-3} \cline{5-7} 
			200000                       & 399004 & 0.0100 &  & 299850 & 599004 & 0.0047 \\ \cline{1-3} \cline{5-7} 
			199976                       & 399004 & 0.0095 &  & 299600 & 599004 & 0.0013 \\ \cline{1-3} \cline{5-7} 
			199850                       & 399004 & 0.0070 &  & 400100 & 799004 & 0.0060 \\ \cline{1-3} \cline{5-7} 
			199720                       & 399004 & 0.0044 &  & 399850 & 799004 & 0.0035 \\ \cline{1-3} \cline{5-7} 
			\multicolumn{1}{|c|}{199600} & 399004 & 0.0020 &  & 399600 & 799004 & 0.0010 \\ \cline{1-3} \cline{5-7} 
			200000                       & 398804 & 0.0120 &  & 199970 & 399204 & 0.0074 \\ \cline{1-3} \cline{5-7} 
			200000                       & 399052 & 0.0095 &  & 200100 & 399350 & 0.0085 \\ \cline{1-3} \cline{5-7} 
			200000                       & 399304 & 0.0070 &  & 300222 & 599444 & 0.0067 \\ \cline{1-3} \cline{5-7} 
			200000                       & 399564 & 0.0044 &  & 300288 & 599444 & 0.0076 \\ \cline{1-3} \cline{5-7} 
			200000                       & 399804 & 0.0020 &  & 379518 & 758546 & 0.0026 \\ \cline{1-3} \cline{5-7} 
		\end{tabular}
		\caption{The table shows the our numerical results of $E^{(0)}$ solved with different parameters $N$ and $n_T$. }
		\label{table: E0}
	\end{table}

    \begin{table}[h]
    	\centering
    	
    	Subleading
    	
    	\begin{tabular}{cccc}
    		\hline
    		\multicolumn{1}{|c|}{energy level} & \multicolumn{1}{c|}{0.0000} & \multicolumn{1}{c|}{0.0100} & \multicolumn{1}{c|}{0.0200}  \\ \hline
    		\multicolumn{1}{|c|}{degeneracy} & \multicolumn{1}{c|}{5472} & \multicolumn{1}{c|}{1283} & \multicolumn{1}{c|}{986}  \\ \hline
    	\end{tabular}
        {\vskip 0.2cm}
    	
    	Exact
    	{\vskip 0.1cm}
    	\begin{tabular}{cccc}
    		\hline
    		\multicolumn{1}{|c|}{energy level} & \multicolumn{1}{c|}{0.0000} & \multicolumn{1}{c|}{0.0100} & \multicolumn{1}{c|}{0.0200}  \\ \hline
    		\multicolumn{1}{|c|}{degeneracy} & \multicolumn{1}{c|}{5475} & \multicolumn{1}{c|}{1289} & \multicolumn{1}{c|}{983}  \\ \hline
    	\end{tabular}
    	\caption{The tables present the degeneracies of dominating energy levels in spectra shown in Figure \ref{fig:subleading-exact-spectrum-nt399004-n200000011}, which are calculated with parameters $n_T=399004$ and $N=200000$. Data for the subleading order action is shown in the table upper, while that for the exact action is shown in the table below.}
    	\label{table: subleading degeneracies}
    \end{table}

    \section{Discussion}\label{discuss}
    
    In this article we have evaluated subleading, in $1/N$, corrections to the dilation operator. This was done by performing and exact analytic evaluation of the one loop mixing between three complex scalar fields, $\phi_1$, $\phi_2$ and $\phi_3$. The operators which mix are constructed using a very large number of $\phi_1$ fields and much fewer $\phi_2$ and $\phi_3$ fields. These operators correspond to excited giant graviton branes, with the $\phi_2,\phi_3$ describing the open string excitations of the branes. The low energy world volume dynamics of the branes is an emergent super Yang-Mills theory. Our computation has allowed us to evaluate ineractions appearing in this emergent gauge theory.
    
    Our exact evaluation gives a detailed formula for the matrix elements of the dilatation operator. This formula has passed a number of nontrivial tests, giving us confidence in the result. First, the terms mixing $\phi_1$ and $\phi_2$ and the terms mixing $\phi_1$ and $\phi_3$ are in complete agreement with the results obtained in \cite{2010Emergentthree,2011Surprisingly}. Secondly, the transformation of the dilatation operator to Gauss graph basis gives formulas in our example that are in complete agreement with the general formulas obtained in \cite{2012Adoublecosetansatz}. Finally, the leading terms mixing $\phi_2$ and $\phi_3$ are in complete agreement with the formulas derived in \cite{2020EmergentYangMills}. By expanding our exact result to the next subleading order, we have obtained the result (\ref{subint}) which is the main result of this paper. 
    
    The formula (\ref{subint}) represent interactions in the emergent Yang-Mills theory. The leading contribution to operator mixing comes from the mixing between $\phi_1$ and $\phi_2$ fields, and the mixing between $\phi_1$ and $\phi_3$ fields. This mixing is diagonalized by the Gauss graph operators. The next correction to operator mixing comes from the mixing between $\phi_2$ and $\phi_3$ fields. In the Gauss graph basis this mixing allows Gauss graphs to mix if they differ in the placement a single closed edge on the graph. The corresponding operators have equal dimensions, so that although this mixes operators with degenerate scaling dimensions, there is no correction to the spectrum of anomalous dimensions. The formula (\ref{subint}) gives a mixing between operators that have different dimension, and consequently it will have a non-trivial effect on the spectrum of operator dimensions.
    
    There are a number of ways in which the study of this article can be extended. In the planar limit, arguments exploiting global symmetries were very helpful in constraining the form of the dilatation operator \cite{2004TheDynamicSpin,2008TheSu(2|2)Dynamic}. Similar computations relevant to our study include \cite{2012FromLargeN,2015RelationBetweenLargeDimension,2014HigherLoopNonplanar,2021OscillatingMultipleGiants,2011NonplanarIntegrability}.
    The studies \cite{2012FromLargeN,2015RelationBetweenLargeDimension} focused on the leading order contribution, while \cite{2014HigherLoopNonplanar,2021OscillatingMultipleGiants,2011NonplanarIntegrability} were focused on the su(2) sector. It would be interesting to see if the result (\ref{subint}) can be recovered by making use of the su(3) symmetry that is present at one loop, and by making use of su(2|2) symmetry at higher loops. For this task, the action of the su(3) generators acting on restricted Schur polynomials \cite{2017RotatingRestrictedSchur} will be a useful result. 
    
    Finally, one question of clear physical significance, is to understand the emergent gauge symmetry. An initial step in this direction was taken in \cite{2020CentralCharges} by showing that the central extension of su(2$\vert$2) generates gauge transformations. This article has given an exact evaluation of the dilatation operator. It would be interesting to explore the arguments of \cite{2020CentralCharges} away from the distant corners approximation, where we expect U(1)$\times$U(1) to be enhanced to U(2). Does the central extension correctly generate gauge transformations in this setting?
	
	\vfill\eject
	
	\appendix
	
	\section{Exact Action of the Dilatation Operator}
	
	In this section, we will respectively show the action of $D_{31},D_{21},D_{32}$ in the restricted Schur polynomials basis. This action is exact, i.e. it is calculated without neither the large $N$ nor the displaced corners limit. The complete result is too large to quote, so we will simply give a few representative examples.
	
	\subsection{Exact Action of $D_{31}$}
	
	$$
	\begin{aligned}
		&DO_{aa}(\xi_0,\xi_1)\\
		=&\frac{12(C_1-3)\xi_0}{(\xi_1+2)(\xi_1+4)(\xi_0+\xi_1+5)}O_{aa}(\xi_0,\xi_1)\\
		-&\frac{2\sqrt{(C_0+1)(C_1-3)}\xi_0}{\xi_1+2}\sqrt{\frac{\xi_1+5}{(\xi_0+1)(\xi_1+3)(\xi_0+\xi_1+5)}}O_{ad}(\xi_0,\xi_1)\\
		+&\frac{2\sqrt{(C_0+1)(C_1-3)}\xi_0(\xi_1-2)}{(\xi_1+2)(\xi_1+4)}\sqrt{\frac{\xi_1+5}{(\xi_0+1)(\xi_1+3)(\xi_0+\xi_1+5)}}O_{ae}(\xi_0,\xi_1)\\
		-&\frac{8(C_1-3)(\xi_1+5)}{(\xi_1+2)(\xi_1+4)^2(\xi_0+\xi_1+5)}\sqrt{\frac{\xi_0(\xi_1+1)(\xi_0+\xi_1+2)}{\xi_1+3}}O_{ea}(\xi_0-1,\xi_1+2)\\
		+&\frac{4\sqrt{(C_0+1)(C_1-3)}}{(\xi_1+3)(\xi_1+4)^2}\sqrt{\frac{\xi_0(\xi_1+1)(\xi_1+5)(\xi_0+\xi_1+2)}{(\xi_0+1)(\xi_0+\xi_1+5)}}O_{ab}(\xi_0-1,\xi_1+2)\\
		+&\frac{2(C_1-3)}{(\xi_1+4)(\xi_0+\xi_1+5)}\sqrt{\frac{\xi_0(\xi_1+1)(\xi_0+\xi_1+2)}{\xi_1+3}}O_{ad}(\xi_0-1,\xi_1+2)\\
		+&\frac{4\sqrt{(C_0+1)(C_1-3)}}{(\xi_1+2)(\xi_1+3)(\xi_1+4)}\sqrt{\frac{\xi_0(\xi_1+1)(\xi_1+5)(\xi_0+\xi_1+2)}{(\xi_0+1)(\xi_0+\xi_1+5)}}O_{ed}(\xi_0-1,\xi_1+2),\\
		-&\frac{2(C_1-3)(\xi_1+6)}{(\xi_1+4)^2(\xi_0+\xi_1+5)}\sqrt{\frac{\xi_0(\xi_1+1)(\xi_0+\xi_1+2)}{\xi_1+3}}O_{ae}(\xi_0-1,\xi_1+2)\\
		+&\frac{4\sqrt{(C_0+1)(C_1-3)}(\xi_1+6)}{(\xi_1+2)(\xi_1+3)(\xi_1+4)^2}\sqrt{\frac{\xi_0(\xi_1+1)(\xi_1+5)(\xi_0+\xi_1+2)}{(\xi_0+1)(\xi_0+\xi_1+5)}}O_{ee}(\xi_0-1,\xi_1+2)
	\end{aligned}
	$$
	
	\vfill\eject
	
	\subsection{Exact Action of $D_{21}$}
	
	$$
	\begin{aligned}
		&DO_{aa}(\xi_0,\xi_1)\\
		=&\frac{12(C_1-3)\xi_0}{(\xi_1+2)(\xi_1+4)(\xi_0+\xi_1+5)}O_{aa}(\xi_0,\xi_1)\\
		-&\frac{2\sqrt{(C_0+1)(C_1-3)}\xi_0}{\xi_1+2}\sqrt{\frac{\xi_1+5}{(\xi_0+1)(\xi_1+3)(\xi_0+\xi_1+5)}}O_{da}(\xi_0,\xi_1)\\
		+&\frac{2\sqrt{(C_0+1)(C_1-3)}\xi_0\xi_1}{(\xi_1+2)^2}\sqrt{\frac{\xi_1+5}{(\xi_0+1)(\xi_1+3)(\xi_0+\xi_1+5)}}O_{ea}(\xi_0,\xi_1)\\
		-&\frac{8\sqrt{(C_1-3)(C_0+1)}\xi_0(\xi_1+1)}{(\xi_1+2)^2(\xi_1+4)}\sqrt{\frac{\xi_1+5}{(\xi_0+1)(\xi_1+3)(\xi_0+\xi_1+5)}}O_{ae}(\xi_0,\xi_1)\\
		+&\frac{4\sqrt{(C_0+1)(C_1-3)}}{(\xi_1+3)(\xi_1+2)^2}\sqrt{\frac{\xi_0(\xi_1+1)(\xi_1+5)(\xi_0+\xi_1+2)}{(\xi_0+1)(\xi_0+\xi_1+5)}}O_{ba}(\xi_0-1,\xi_1+2)\\
		+&\frac{2(C_1-3)}{(\xi_1+4)(\xi_0+\xi_1+5)}\sqrt{\frac{\xi_0(\xi_1+1)(\xi_0+\xi_1+2)}{\xi_1+3}}O_{da}(\xi_0-1,\xi_1+2)\\
		-&\frac{2(C_1-3)(\xi_1+8)}{(\xi_1+2)(\xi_1+4)(\xi_0+\xi_1+5)}\sqrt{\frac{\xi_0(\xi_1+1)(\xi_0+\xi_1+2)}{\xi_1+3}}O_{ea}(\xi_0-1,\xi_1+2)\\
		+&\frac{4\sqrt{(C_0+1)(C_1-3)}}{(\xi_1+2)(\xi_1+3)(\xi_1+4)}\sqrt{\frac{\xi_0(\xi_1+1)(\xi_1+5)(\xi_0+\xi_1+2)}{(\xi_0+1)(\xi_0+\xi_1+5)}}O_{de}(\xi_0-1,\xi_1+2),\\
		+&\frac{4\sqrt{(C_0+1)(C_1-3)}\xi_1}{(\xi_1+2)^2(\xi_1+3)(\xi_1+4)}\sqrt{\frac{\xi_0(\xi_1+1)(\xi_1+5)(\xi_0+\xi_1+2)}{(\xi_0+1)(\xi_0+\xi_1+5)}}O_{ee}(\xi_0-1,\xi_1+2)
	\end{aligned}
	$$
	
	\vfill\eject
	
	\subsection{Exact Action of $D_{32}$}
	
	$$
	DO_{aa}(\xi_0,\xi_1)=0
	$$
	
	$$
	\begin{aligned}
		&DO_{ba}(\xi_0,\xi_1)\\
		=&\frac{8(C_1-1)(\xi_1-1)}{\xi_1^3(\xi_0+\xi_1+3)}O_{ba}(\xi_0,\xi_1)\\
		+&\frac{4\sqrt{(C_1-1)(C_0-1)}}{\xi_1^2}\sqrt{\frac{\xi_1-1}{(\xi_0+3)(\xi_1+1)(\xi_0+\xi_1+3)}}O_{db}(\xi_0,\xi_1)\\
		+&\frac{4\sqrt{(C_1-1)(C_0-1)}(\xi_1+2)}{\xi_1^3}\sqrt{\frac{\xi_1-1}{(\xi_0+3)(\xi_1+1)(\xi_0+\xi_1+3)}}O_{eb}(\xi_0,\xi_1)\\
		-&\frac{4\sqrt{(C_1-1)(C_0-1)}}{\xi_1^2}\sqrt{\frac{(\xi_1-1)(\xi_1+1)}{(\xi_0+3)(\xi_0+\xi_1+3)}}O_{bd}(\xi_0,\xi_1)\\
		+&\frac{2(C_1-1)(\xi_1-1)}{\xi_1(\xi_1+1)(\xi_0+\xi_1+3)}O_{dd}(\xi_0,\xi_1)\\
		+&\frac{2(C_1-1)(\xi_1-1)(\xi_1+2)}{\xi_1^2(\xi_1+1)(\xi_0+\xi_1+3)}O_{ed}(\xi_0,\xi_1)\\
		+&\frac{4\sqrt{(C_1-1)(C_0-1)}(\xi_1-2)}{\xi_1^3}\sqrt{\frac{(\xi_1-1)(\xi_1+1)}{(\xi_0+3)(\xi_0+\xi_1+3)}}O_{be}(\xi_0,\xi_1)\\
		-&\frac{2(C_1-1)(\xi_1-1)(\xi_1+2)}{\xi_1^2(\xi_1+1)(\xi_0+\xi_1+3)}O_{de}(\xi_0,\xi_1)\\
		-&\frac{2(C_1-1)(\xi_1-1)(\xi_1+2)^2}{\xi_1^3(\xi_1+1)(\xi_0+\xi_1+3)}O_{ee}(\xi_0,\xi_1)
	\end{aligned}
	$$

	\acknowledgments
	
	The author sincerely appreciates the instruction from his supervisor, Prof. Robert de Mello Koch. This work can not be finished without his help. The author also thanks Dr. Lei Yin for useful discussions. This work is supported by the Guangdong Major Project of Basic and Applied Basic Research No. 2020B0301030008 and the National Natural Science Foundation of China under Grant No. 12035007.

	
	\bibliographystyle{JHEP}
	\bibliography{reference2021_12_16}
	
\end{document}